\documentclass[a4paper,11pt]{article}
\pdfoutput=1 

\usepackage{jheppub} 

\usepackage{caption} 
\usepackage[]{color}
\renewcommand{\thesection}{\arabic{section}}

\title{\boldmath Hairy Black Holes in AdS$_5\times S^5$}

 \author{Julija Markeviciute, }
 \author{Jorge E. Santos}
 \affiliation{Department of Applied Mathematics and Theoretical Physics, \\
 University of Cambridge, \\ Cambridge, CB3 0WA, UK}
 \emailAdd{j.markeviciute}
 \emailAdd{j.e.santos@damtp.cam.ac.uk}
\abstract{We use numerical methods to exhaustively study a novel family of hairy black hole solutions in AdS$_5$. These solutions can be uplifted to solutions of type IIB supergravity with AdS$_5\times S^5$ asymptotics and are thus expected to play an important role in our understanding of AdS/CFT. We find an intricate phase diagram, with the aforementioned family of hairy black hole solutions branching from the Reissner-Nordstr\"om black hole at the onset of the superradiance instability. We analyse black holes with spherical and planar horizon topology and explain how they connect in the phase diagram. Finally, we detail their global and local thermodynamic stability across several ensembles. }
\begin{document} 
\maketitle
\flushbottom
\newpage
 
\section{\label{sec:introduction}Introduction}
Black holes play a paramount role in gauge-gravity dualities. In particular, they are typically the dominant saddle points in the high temperature regime of the canonical ensemble. One can thus gain insight into typical states of conformal field theories at high temperature by studying novel black hole solutions that are asymptotically anti-de Sitter (AdS).

In this paper we will devote our attention to a particular conformal field theory (CFT) living on the Einstein static universe $\mathbb{R}_t\times S^3$, namely $\mathcal{N}=4$ SYM with gauge group $SU(N)$. There are many reasons to study this particular CFT, perhaps the most important being that this is the theory for which AdS/CFT was first formulated in \cite{Maldacena:1997re}, and for which our holographic dictionary is best understood \cite{Gubser:1998bc,Witten:1998qj,Aharony:1999ti}.

In \cite{Maldacena:1997re}, the strong coupling limit of $\mathcal{N}=4$ SYM at large t'Hooft coupling and at infinite gauge group rank $N$ was conjectured to be IIB supergravity on AdS$_5\times S^5$. We are thus led to consider black hole solutions of IIB supergravity with AdS$_5\times S^5$ asymptotics if we want to understand the thermodynamic saddle points of $\mathcal{N}=4$ SYM living on the Einstein static universe. Since we will be working in the supergravity limit, we will be considering states on the CFT at energies of order $N^2$.

However, even studying black hole solutions in IIB supergravity is far from being an easy task. For instance, small Schwarzschild-AdS black holes can be shown to be unstable to a localisation on the $S^5$ if their radius in AdS units is sufficiently small \cite{Banks:1998dd,Peet:1998cr,Hubeny:2002xn,Buchel:2015gxa}. The bumpy black holes in AdS$_5\times S^5$ that branch from the onset of this instability were only recently constructed and necessarily require solving partial differential equations \cite{Dias:2015pda}. The work developed in this paper ignores such instabilities and focuses on solutions of five-dimensional $\mathcal{N}=8$ gauged supergravity, which is thought to be a consistent truncation of IIB supergravity on AdS$_5\times S^5$ \footnote{This has actually never been shown in full generality, partially because of the self dual condition imposed on the Ramond-Ramond $F_5$ form flux, even though interesting progress has been recently made in \cite{Ciceri:2014wya}.}. This truncation is such that enough symmetry is assumed so that the localisation phenomenon described above does not occur.

Even dealing with all the supergravity fields of five-dimensional $\mathcal{N}=8$ gauged supergravity proves to be a rather complicated task. In order to bypass this, we will focus on a truncation of $\mathcal{N}=8$ supergravity which, to our knowledge, was first proposed in \cite{Bhattacharyya:2010yg}. The spectrum of five-dimensional gauged $\mathcal{N}=8$ supergravity comprises one graviton, 42 scalars, 15 gauge fields and 12 form fields. The consistent truncation that we are going to consider contains the graviton, a complex scalar field and a Maxwell field, under which the scalar field is charged. For more details on this truncation we refer the reader to \cite{Bhattacharyya:2010yg}. Once the dust settles, the action reads:

\begin{multline}
\label{eq:action}
 S=\frac{1}{16\pi G_5}\int \mathrm{d}^5x\sqrt{g}\left\{R[g]+12-\frac{3}{4}F_{\mu\nu}F^{\mu\nu}-\frac{3}{8}\left[(D_\mu\phi)(D^\mu\phi)^\dagger-\frac{\nabla_\mu\lambda\,\nabla^\mu\lambda}{4(4+\lambda)}-4\lambda\right]\right\}
 \\
 -\frac{1}{16\pi G_5}\int \mathrm{d}^5x\, F\wedge F\wedge A,
\end{multline} 
where $F_{\mu\nu}=2\partial_{[\mu}A_{\nu]}$, $D_\mu\phi=\nabla_\mu\phi-i\,e A_\mu\phi$, $e=2$, $\lambda = \phi \phi^\dagger$, the radius of AdS$_5$ is set to unity and $G_5=\pi/(2N^2)$ is the five-dimensional Newton's constant. We note that at this stage we have already used AdS/CFT, in the sense that $G_5$ is given in terms of the rank of the gauge group of $\mathcal{N}=4$ SYM. The tachyonic scalar field $\phi$ has the charge $e=2$ and $m_\phi^2=-4$, which saturates the five-dimensional Breitenl\"ohner-Freedman (BF) bound \cite{Breitenlohner:1982jf}.

The couplings and scalar field charges that come from this embedding in IIB have very particular forms and values. Indeed, in \cite{Basu:2010uz,Dias:2011tj} a bottom up model that shares many features with (\ref{eq:action}) was considered. There, the scalar field charge $e$ was a free parameter and the self-coupling potential of the scalar field was a simple mass term - the action was that of the Abelian Higgs model in AdS. The details of the phase diagram will turn out to depend rather nontrivially on the specific form of the action (\ref{eq:action}). Nevertheless, the authors of \cite{Basu:2010uz,Dias:2011tj} concluded that small black holes in AdS are afflicted by an instability, so long as the charged scalar field has sufficiently large $e$. The origin of this instability goes back to the so-called superradiant scattering \cite{Starobinsky:1973}, which can only occur for charged scalar field satisfying $\omega< e\,\mu$, where $\omega$ is the frequency used in the scattering process and $\mu$ the chemical potential of the background solution.

Interestingly enough, from the analysis of \cite{Basu:2010uz,Dias:2011tj}, it was not clear whether scalar fields with charge $e=2$ could be unstable to the superradiant instability. The reason for this is worth emphasising: normal modes of scalar fields saturating the BF bound around pure AdS have an energy gap given by $\omega = 2$. Furthermore, we will see that charged black holes maximise $\mu$ at fixed energy when they are extremal. Finally, if the black hole is small, one can show that $\mu\simeq 1$ for extremal holes. If we now use our condition for superradiant scattering, one concludes that the system will be unstable if $e> 2$. We note that this does not mean that \emph{other} types of instabilities cannot exist even for small values of the scalar field charge\footnote{For instance, large extremal black holes are known to be unstable against neutral scalar field perturbations, but the triggering mechanism for this instability is \emph{not} superradiance.}. We shall see that the superradiant instability is pervasive even for small charged black hole solutions of (\ref{eq:action}).

There are a handful of solutions to the equations of motion derived from (\ref{eq:action}) that are known to be analytic, the most general being the Kerr-Reissner-Nordstr\"om black hole \cite{Gibbons:2004uw,Gibbons:2004ai,Cvetic:2004ny,Chong:2005hr,Chong:2005da,Chong:2006zx,Cvetic:2005zi,Gutowski:2004yv,Gutowski:2004ez,Kunduri:2006ek,Wu:2011gq}. They all have one thing in common, namely that the charged scalar field vanishes. Hairy solutions, \emph{i.e.} solutions with a nontrivial scalar field profile $\phi$, were first constructed in a matched asymptotic expansion in \cite{Bhattacharyya:2010yg}, where the black holes were taken to be arbitrarily small. In this paper we construct the novel hairy black hole solutions of (\ref{eq:action}) at the full nonlinear level, \emph{i.e.} our black holes are not necessarily small. As a test of our numerical procedure, we give a detailed comparison with the analytic results of \cite{Bhattacharyya:2010yg}.
 
\begin{figure}[!htpb]
\centering
    \begin{minipage}[t]{0.45\textwidth}
    \includegraphics[width=\textwidth]{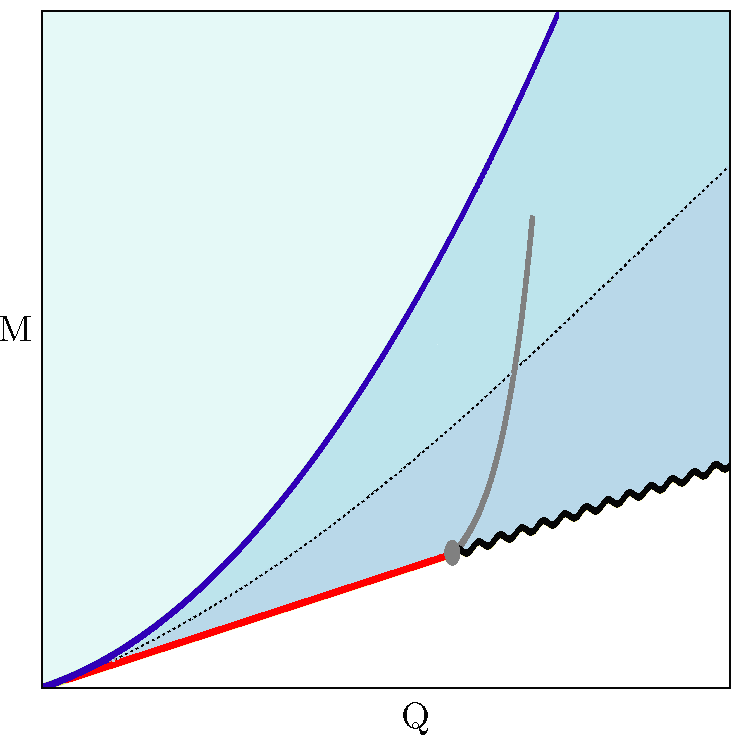}
  \end{minipage}
  \caption{\label{fig:eg}The proposed microcanonical phase diagram by Bhattacharyya \emph{et al.} (taken from \cite{Bhattacharyya:2010yg}, not drawn to scale). The lower solid line is the BPS bound on which the supersymmetric soliton resides. The straight red segment represents smooth branch and the wiggly black part represents singular soliton. Hairy black holes were proposed to exist between the curve indicating the onset of the superradiant instability (solid blue) and the BPS bound. The dotted black curve is the extremal RNAdS black holes. The grey solid line shows a possible phase transition between two different types of hairy black holes, with different zero size limits.}
\end{figure}

Our findings will be consistent with those of \cite{Bhattacharyya:2010yg} for sufficiently small asymptotic charges. We thus start by reviewing their conjectured phase diagram, which is depicted in Fig.~\ref{fig:eg}. The perturbative analysis carried by Bhattacharyya \emph{et al.} shows that in the phase diagram infinitesimally small hairy black holes smoothly join to a horizonless solitonic solution saturating the five dimensional BPS bound (red solid line in Fig.~\ref{fig:eg}). Note that because of the unusual normalisation of the kinetic term for the photons in (\ref{eq:action}) the supersymmetric bound occurs for solutions satisfying $M=3|Q|$. This supersymmetric soliton was then numerically constructed for large values of the charge $|Q|$, and was found to become singular at a specific charge $Q_c$. The approach to this critical charge revealed an intricate spiralling behaviour. Bhattacharyya \emph{et al.} went further, and constructed a singular solitonic solution that extended to infinite values of $Q$, and approached the same spiral as the regular soliton, but from values above $Q_c$ (wiggly black line in Fig.~\ref{fig:eg}).

In \cite{Bhattacharyya:2010yg} a number of possibilities were envisaged for the behaviour of large black holes in this system. We aim to finally unravel the full nonlinear picture. This paper is organised as follows: section \ref{sec:setup} introduces in more detail the setup that we are considering, including the equations of motion derived from (\ref{eq:action}). In section \ref{sec:num}, we detail the numerical method we used to solve this problem. In section \ref{sec:results}, we present our main results; in section \ref{sec:comparison}, we compare the full nonlinear results to those obtained in \cite{Bhattacharyya:2010yg} and section \ref{sec:discussion} concludes the paper with a discussion and future directions. 

\section{\label{sec:setup}Setup}

The consistent truncation is described by the five-dimensional Einstein-Maxwell AdS gravity coupled to a charged complex scalar field with action given as in (\ref{eq:action}). The equations of motion derived from \eqref{eq:action} are
\begin{subequations}
\label{eq:a}
\begin{align}
\label{eq:a:eeq}
&G_{\mu\nu}-6 g_{\mu\nu}=\frac{3}{2}T_{\mu\nu}^{EM}+\frac{3}{8}T_{\mu\nu}^{mat}\\
\label{eq:a:maxwell}
&\nabla_\lambda F_\mu{}^\lambda=\frac{1}{4}\varepsilon^{\mu\nu\rho\alpha\lambda}F_{\mu\nu}F_{\rho\alpha}+\frac{i}{4}\left[\phi(D_\mu\phi)^\dagger-\phi^\dagger D_\mu\phi\right]\\
\label{eq:a:scalar}
&D_\mu D^\mu\phi+\phi\left[\frac{(\nabla_\mu\lambda)(\nabla^\mu\lambda)}{4(4+\lambda)^2}-\frac{\nabla_\mu \nabla^\mu \lambda}{2(4+\lambda)}+4\right]=0,
\end{align}
\end{subequations} 
where
\begin{align*}
T_{\mu\nu}^{EM}&=F_\mu{}^\lambda F_{\nu\lambda}+\frac{1}{4}g_{\mu\nu}\,F^2\\
T_{\mu\nu}^{mat}&=\frac{1}{2}\left[D_\mu\phi\,(D_\nu\phi)^\dagger+D_\nu\phi\,(D_\mu\phi)^\dagger\right]-\frac{1}{2}g_{\mu\nu}(D_\alpha\phi)(D^\alpha\phi)^\dagger+2g_{\mu\nu}\, \lambda\\
&-\frac{1}{4(4+\lambda)}\left[(\nabla_\mu\lambda)(\nabla_\nu\lambda)-\frac{1}{2}g_{\mu\nu}(\nabla_\alpha\lambda)(\nabla^\alpha\lambda)\right].
\end{align*}
We look for static, spherically symmetric and asymptotically global AdS$_5$ solutions and for now we will not specify our gauge choice\footnote{Note that we have fixed the $U(1)$ gauge freedom by taking $\phi$ to be real.}:
\begin{equation}
 \mathrm{d}s^2=-f(r)\mathrm{d}t^2+g(r)\mathrm{d}r^2+\Sigma(r)^2\mathrm{d}\Omega^2_3,\quad A_\mu\mathrm{d}x^\mu=A(r)\mathrm{d}t,\quad \phi=\phi^\dagger=\phi(r).
\end{equation}
Since our solutions are only electrically charged, the Chern-Simons term (first term on the left hand side of Eq.~(\ref{eq:a:maxwell})) plays no role. The Einstein equation, the Maxwell equation and and the scalar equation \eqref{eq:a} yield a system of four equations \cite{Bhattacharyya:2010yg}:
\begin{align}
\label{eq:eom}
\begin{split}
f'&-\frac{f}{2 \Sigma \Sigma'}\left[4 g+g\Sigma^2 \left(8+\phi^2\right)-4 \Sigma'^2+\frac{\Sigma^2 \phi'^2}{4+\phi^2}\right]-\frac{\Sigma}{2 \Sigma'}\left(A^2\phi^2g-A'^2\right)=0, \\
g'&+g^2 \left(\frac{4}{\Sigma \Sigma'}+\frac{8 \Sigma}{\Sigma'}+\frac{\Sigma\phi^2}{\Sigma'}\right)+g\left(\frac{f'}{f}+\frac{\Sigma A'^2}{f\Sigma'}+\frac{4 \Sigma'}{\Sigma}+\frac{2 \Sigma''}{\Sigma'}\right)=0, \\
A''&+\frac{1}{2}\left(\frac{6\Sigma'}{\Sigma}-\frac{f'}{f}-\frac{g'}{g}\right)A'-g\,\phi^2A=0, \\
\phi''&+\frac{1}{2}\left(\frac{6 \Sigma'}{\Sigma}+\frac{f'}{f}-\frac{g'}{g}\right)\phi'-\frac{\phi}{4+\phi^2}\,\phi'^2+\frac{g}{f}\left(A^2+f\right)\left(4+\phi^2\right)\phi=0.
\end{split}
\end{align}
The $'$ denotes the derivative with respect to $r$.

At this point we pick a gauge where $\Sigma(r)=r$, so that $r$ measures the radius of the round $S^3$ in AdS$_5$. In this gauge, we require that our solutions are asymptotically AdS$_5$, \emph{i.e.} at large $r$ they must satisfy the following expansion \cite{Ashtekar:1984zz,Henneaux:1985tv,Henningson:1998gx,deHaro:2000vlm}
\begin{align}
\begin{split}
f(r)&=r^2+1+\mathcal{O}(r^{-2}),\quad g(r)=\frac{1}{1+r^2}+\mathcal{O}(r^{-6}),\\
A(r)&=\mu+\mathcal{O}(r^{-2}),\quad\phi(r)=\frac{\varepsilon}{r^2}+V\frac{\log r}{r^2}+\mathcal{O}(r^{-4})\,,
\end{split}
\label{eq:expansion}
\end{align}
where $\mu$ is the chemical potential and the constants $V$ and $\varepsilon$ will shortly be identified. Using the AdS/CFT correspondence \cite{Gubser:1998bc,Witten:1998qj}, $V$ is regarded as the source for the operator dual to $\phi$ and $\varepsilon$ is its expectation value, \emph{i.e.} $\varepsilon = \langle \mathcal{O}_\phi \rangle$. This choice implicitly assumes standard quantisation. The operator dual to $\phi$ has conformal scaling dimension $\Delta=2$. We will be interested in solutions representing states of the conformal field theory that are not sourced, so we will set $V=0$. These normalizable conditions give rise to a four parameter set of asymptotically AdS$_5$ solutions to~\eqref{eq:eom} \cite{Bhattacharyya:2010yg}. Further imposing suitable regularity and normalisability conditions results in two parameter space of solutions which may be taken to be the mass ($M$) and charge $(Q)$ of the black hole, with $\varepsilon$ and $\mu$ being determined as a function of $M$ and $Q$.

The frequency of the lowest normal mode of $\phi$ is $\Delta = 2$. In \cite{Basu:2010uz} it was shown that small Reissner-Nordstr\"om AdS (RNAdS) black holes suffer from superradiant instability whenever $e\mu>\Delta$, where $\mu$ is a chemical potential of the black hole. For RNAdS black holes $\mu\leq (1+2R^2)$, where $R$ is the Schwarzschild radius of the black hole\footnote{Defined so that the entropy for the RNAdS BH is $S=\pi R^3$.}, therefore, small black holes satisfy $\mu\leq 1$ (saturating at extremality). Hence small charged black holes are always stable when $e<e_c=\Delta$ and in our setup small near extremal black holes lie at the edge of the instability. These small near extremal charged black holes are unstable to the superradiant tachyon condensation and evolve towards a small black hole with the charged scalar hair. 

\subsection{Known solutions}
All known solutions are found in the radial gauge where $\Sigma(r)=r$.
\subsubsection{The Reissner-Nordstr\"om black hole}
If we switch off the scalar field we recover the familiar Reissner-Nordstr\"om two parameter set of solutions to~\eqref{eq:eom}
\begin{align}
\label{eq:RN} 	
\begin{split}
&f(r)=\frac{\mu^2 R^4}{r^4}-\frac{(R^2+\mu^2+1)R^2}{r^2}+r^2+1,\\
&g(r)=\frac{1}{f(r)},\quad A(r)=\mu\left(1-\frac{R^2}{r^2}\right),\quad\phi(r)=0.
\end{split}
\end{align}
We record the thermodynamic formulae for later use (henceforth all the thermodynamic quantities will be scaled by N$^2$)
\begin{align}
\begin{split}
M&=\frac{3}{4}R^2\left(1+R^2+\mu^2\right)\\
Q&=\frac{1}{2}\mu R^2\\ 
S&=\pi R^3\\
T&=\frac{1}{2\pi R}\left(1+2R^2-\mu^2 \right).
\end{split}
\end{align}
Note that $R$ is the outer horizon if the condition $\mu^2\leq(1+2R^2)$ is satisfied. This inequality is saturated at extremality, where $T=0$. The resulting extremal black hole is regular, and has a degenerate bifurcating Killing horizon.
 
\subsubsection{The BPS solitons}
In this section we briefly outline the numerical study of the spherically symmetric smooth and singular solitons given in \cite{Bhattacharyya:2010yg}. We shall see later that these can be regarded as the BPS limit of the hairy black hole configurations. Soliton solutions are easier to determine, since they are known to be supersymmetric. Instead of solving the equations of motion (\ref{eq:eom}) directly one resorts to searching for nontrivial solutions of the Killing spinor equations, which are first order in space. After some nontrivial manipulations, one can cast \emph{any} supersymmetric solution of the action (\ref{eq:action}) into the following form
\begin{multline}
f(r)=\frac{1+\rho^2h^3}{h^2}\,,\qquad g(r)=\frac{4\rho^2h^2}{(2\rho h+\rho^2\dot{h})^2(1+\rho^2h^3)}
\\
A(r)=\frac{1}{h(r)}\,,\qquad \phi(r)=2\left[\left(h+\frac{\rho \dot{h}}{2}\right)^2-1\right]^{1/2} 
\end{multline}
where the $\dot{\,}$ denotes the derivative with respect to the variable $\rho$, given by $r^2=\rho^2h$, and $h$ has to satisfy the following second order differential equation
\begin{equation}
\rho\left(1+\rho^2h^3\right)\ddot{h}+\left(3+7\rho^2h^3+\rho^3h^2\dot{h}\right)\dot{h}-4\rho\left(1-h^2\right)h^2=0.
\label{eq:soliton}
\end{equation}
This equation has a number of remarkable properties. Perhaps the most striking being that at large $\rho$ it demands $h(\rho)|_{\rho\rightarrow\infty}=1$. This condition automatically ensures normalisability of the physical fields $f$, $g$, $A$ and $\phi$. At the origin, $r=0$ or equivalently $\rho=0$, there are a number of possibilities. Assuming that solutions to (\ref{eq:soliton}) behave as
\begin{equation}
\lim_{\rho\to0}h = \frac{h_\alpha}{\rho^\alpha}\,, 
\end{equation}
gives the following possible exponents $\alpha=0,1,2/3,2$. Solutions with $\alpha = 0$ are regular, and the remaining are singular. For each of these exponents we can find solitonic solutions, but the dimension of their moduli space strongly depends on $\alpha$. For $\alpha=0,1$ there is a one parameter family of solutions, for $\alpha=2/3$ there is a unique solution and for $\alpha =2$ the solution spans a two dimensional moduli space. In addition, in \cite{Bhattacharyya:2010yg} it was shown that the smooth soliton solutions with $\alpha=0$ exist for small values of the charge $Q$ and that the singular solitonic solution with $\alpha = 1$ exists for large values of $Q$. The two families merge precisely at a special point which is given by the singular soliton with $\alpha = 2/3$.

\begin{figure}[t]
\centering
  \begin{minipage}[t]{0.45\textwidth}
    \includegraphics[width=\textwidth]{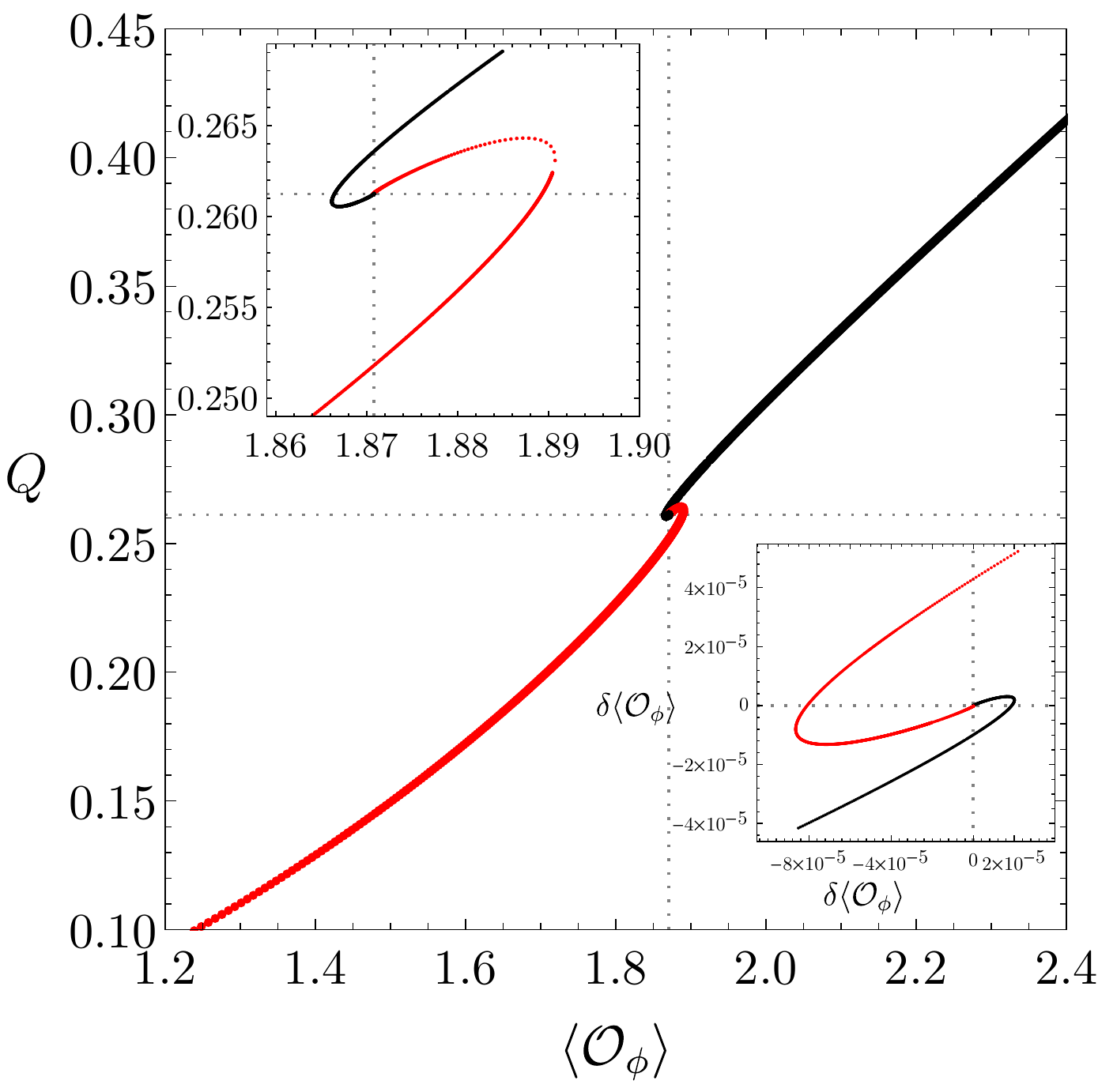}
  \end{minipage}
 \hfill
    \begin{minipage}[t]{0.45\textwidth}
    \includegraphics[width=\textwidth]{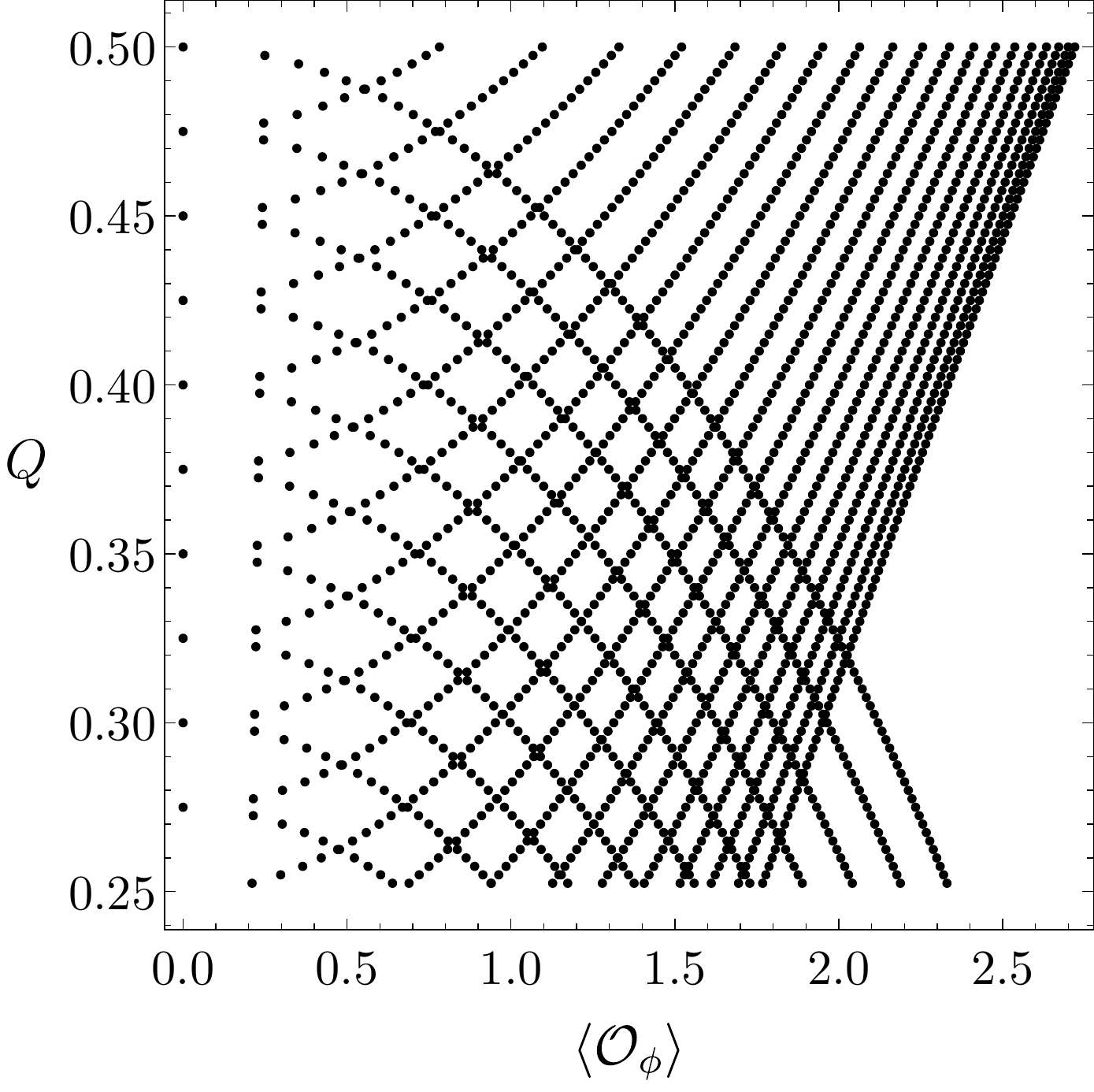} 
  \end{minipage}
      \caption{\textit{Left}: Charge of the solitonic solutions $Q$ versus the vacuum expectation value of the dual operator $\langle\mathcal{O_\phi}\rangle$. The black solid line from above is the singular soliton and the red line from below is the smooth solution. The dotted gridlines show coordinates of the special solution with $\alpha=2/3$. \textit{Right:} The $\alpha=2$ soliton solutions. The wedges are for constant $h_2$ which decreases from $1$. These solution curves appear to extend to $|Q|\rightarrow+\infty$. We also did not find any limiting value for $\langle\mathcal{O_\phi}\rangle$.}
        \label{fig:solitons}
\end{figure}

We reproduce the results of \cite{Bhattacharyya:2010yg}. The line of smooth solitons terminates at the singular solution with the ``critical'' value $Q_c\simeq0.2613$ as the central density $h_0\rightarrow\infty$; the family of singular solitons with $\alpha=1$ branches out of this point at $h_1\rightarrow 0$, extending to higher charges. The critical charge $Q_c$ can also be obtained by solving for the solution with $\alpha = 2/3$ and $h_{2/3}=1$, thus confirming the picture of \cite{Bhattacharyya:2010yg}. In addition, Bhattacharyya \emph{et al.} analysed the asymptotic behaviour around the limiting solution analytically and proposed that these two soliton branches exhibit damped (possibly periodic) oscillations around $Q_c$ in the space parametrized by $Q$ and $\left<\mathcal{O}_\phi\right>$, resulting in an infinite discrete non-uniqueness of the soliton solutions as $Q\rightarrow Q_c$ (see Fig.~\ref{fig:solitons}). Note that there exists a maximum charge $Q_{max}\simeq 0.2643$ for the $\alpha=0$ family and a minimum charge $Q_{min}\simeq 0.2605$ for the $\alpha=1$ singular soliton. The limiting expectation value for the operator dual to the scalar field is $\left<\mathcal{O}_\phi\right>_c\sim 1.8710$, and corresponds to the $\alpha=2/3$ singular solution.

We also compute the singular $\alpha=2$ case which provides a two parameter class of solutions parametrized by $h_2$ and $\left.\rho^3 \partial_{\rho}h\right|_{\rho\rightarrow\infty}$ \footnote{In this case we use the same numerical method as for the hairy black holes, instead of solving (\ref{eq:soliton}) directly. We will detail the numerical method shortly.}. The latter can be regarded as setting the charge $Q$, therefore, for any $h_2$, solutions exist with any value of the charge. It appears that these solutions are not connected to the other solutions studied in this paper (see Fig.~\ref{fig:solitons}).
 
\section{\label{sec:num}Numerical Construction of Hairy Black holes}

We use the DeTurck method \cite{Headrick:2009pv} (for an extensive review see \cite{Dias:2015nua}) which allows us to instead solve the Einstein-DeTurck or harmonic Einstein equation
\begin{equation}
\label{eq:deturck}
G_{\mu\nu}-\nabla_{(\mu}\xi_{\nu)}=0, 
\end{equation}
where $\xi^{\mu}=g^{\nu\rho}\left[\Gamma^\mu_{\nu\rho}(g)-\Gamma^\mu_{\nu\rho}(\tilde{g})\right]$ is the DeTurck vector and $\tilde{g}$ is a reference metric of our choice such that it possess the same causal structure of our desired solution $g$. This method is very useful because as we solve~\eqref{eq:deturck} the gauge is automatically fixed by the condition ${\xi^{\mu}=0}$. Static solutions to the harmonic Einstein equation under certain regularity assumptions will also satisfy the Einstein equation \cite{Figueras:2011va}. However, in this case we do not know whether there exist solutions with $\xi^{\mu}\ne 0$ (so called Ricci solitons). We check \textit{a posteriori} that the solutions presented in this paper satisfy $\xi^{\mu}=0$ at least to $\mathcal{O}(10^{-10})$ precision and also demonstrate good convergence (see Appendix~\ref{sec:numval}, Figs.~\ref{fig:deturck}-\ref{fig:planarconv}). To solve~\eqref{eq:deturck} we use Newton-Raphson method with pseudospectral collocation on a Chebyshev grid to discretise the equations.

We make a compact coordinate change $r=\dfrac{y_+}{\sqrt{1-y^2}}$ so that $y=1$ corresponds to $r=\infty$ and $y=0$ to $r=y_+$. The first metric \emph{ansatz} that we use is
\begin{align}
\label{eq:metric1}
\mathrm{d}s^2_1&=\frac{1}{1-y^2}\left[-y^2\Delta(y)q_1 \mathrm{d}t^2+\frac{y_+^2 q_2 \mathrm{d}y^2}{\left(1-y^2\right)\Delta(y)}+y_+^2 q_3\mathrm{d}\Omega_3^2\right]
\end{align}
together with
\begin{multline}
A(r)=y^2q_4(y)\,,\quad \phi(r)=\left(1-y^2\right)q_5(y)\quad \\ \text{and}\quad \Delta(y)=1+ 2y_+^2-\tilde{\mu} ^2-\left(1+y_+ ^2-2 \tilde{\mu} ^2\right)y^2-\tilde{\mu} ^2 y^4\,.
\label{eq:ansatz1gauge}
\end{multline}
Our reference metric $\tilde{g}$ used in the DeTurck method is obtained from (\ref{eq:metric1}) by setting $q_1=q_2=q_3=1$. This is simply the metric of a RNAdS when $\tilde{\mu}=\mu$ \eqref{eq:RN}. The parameter $\tilde{\mu}$ is left to be specified freely as it just sets the reference metric and is in general  different from the chemical potential of the physical metric.

As we want to explore the solution space we start somewhere on the merger line, \emph{i.e.} on a solution with some parameter coordinates ($\mu$, $y_+$) for which $\phi$ is arbitrarily small. If we want to probe low temperatures a natural choice is $\mu=1$ (as black holes with $y_+<1/2$ are small and have $\mu\sim 1$). The physical chemical potential of the black hole is then given by the gauge field on the boundary, $\mu=A(r)|_{r\to \infty}=q_4(1)$ (see the expansion (\ref{eq:expansion})).

At the conformal boundary, located at $y=1$, we demand that ${q_1(1)=q_2(1)=q_3(1)=1}$, $q_5(1)=\epsilon$ and a Robin condition $\left[y_+^2q_5'-2q_4^2q_5\right]_{y=1}=0$ for the gauge field which ensures that Newton's method converges to the hairy solution if we specify nonzero $\epsilon$. Note that $\epsilon$ is related to $\varepsilon$ via $\epsilon = y_+^2 \varepsilon$.
 
Regularity at the horizon demands $q_1(0)=q_2(0)$ and pure Neumann for the remaining functions, \emph{i.e.} $q_i'(0)=0$. In many regions of the parameter space, $\epsilon$ will not uniquely parametrise a solution, however the strength of the scalar field at the horizon, $q_5(0)\equiv\epsilon_0$, will. Depending on which region of parameter space we want to probe, we might decide to parametrise our solution with $\epsilon$ or $\epsilon_0$. Thus we are left with two parameters after we fix $\tilde{\mu}$, namely $(y_+,\epsilon)$ or $(y_+,\epsilon_0)$.

However, the RNAdS-like \emph{ansatz} (\ref{eq:metric1}) does not have good convergence properties almost everywhere in moduli space. We found that the following \emph{ansatz} has better convergence properties (at least a few order of magnitudes better!) if we simply set $\Delta(y)=y_+^2$ in (\ref{eq:metric1}), yielding:
\begin{align}
\label{eq:metric2}
\mathrm{d}s^2_2&=\frac{1}{1-y^2}\left[-y^2y_+^2q_1 \mathrm{d}t^2+\frac{q_2 \mathrm{d}y^2}{\left(1-y^2\right)}+y_+^2 q_3\mathrm{d}\Omega_3^2\right].
\end{align}
The trade off is that now the functions at high central field density $\epsilon_0$ are more peaked at high temperatures, therefore we use the \emph{ansatz} \eqref{eq:metric1} to extend our solution curves in the high $\epsilon_0$, high $T$ regime. The boundary conditions remain the same except for the gauge field, which in the new \emph{ansatz} obeys to the following boundary condition $\left[y_+^2(q_5+q_5')+q_5-2q_4^2q_5\right]_{y=1}=0$. This boundary condition can be obtained by solving the Einstein-DeTurck equations near the boundary.

It is not always easy to find a reference metric for the DeTurck method, but here we have the luxury of having two good reference metrics. The results obtained with the two different reference metrics match at least to $0.1\%$ numerical accuracy in all the physical quantities such as energy (for the quantitative comparison of the two \emph{ansatz} see Fig.~\ref{fig:ansatz}, Appendix~\ref{sec:numval}).

We present thermodynamic formulae for the line element \eqref{eq:metric2} since this was the \emph{ansatz} we used the most. The electric charge is obtained by computing the flux of the electromagnetic field tensor at infinity
\begin{equation}
Q=\dfrac{1}{4}A'(r)|_{r\to \infty}=\frac{y_+^2}{4}\left.\left(2q_4+\frac{\mathrm{d} q_4}{\mathrm{d}y}\right)\right|_{y=1}.
\end{equation}

We compute the Hawking temperature of the black hole by requiring smoothness of the Euclidean spacetime and it is simply given by
\begin{equation}
T=\frac{y_+}{2\pi}\,.
\end{equation}
The entropy of a BH is proportional to its horizon area and is given by
\begin{equation}
S=\pi y_+^3\,q_3(0)^{3/2}\,.
\end{equation}
 
To compute the mass of the black hole we use the Ashtekar-Das formalism \cite{Ashtekar:1999jx}
\begin{equation}
M=\frac{y_+^2}{8}\left[1+3y_+^2+y_+^4\left(2-q_5^2-q_1''\right)\right]_{y=1}.
\end{equation}
We further checked that this matched the holographic renormalization technique of \cite{Henningson:1998gx,Balasubramanian:1999re,deHaro:2000vlm} up to the energy of the ground state of the global AdS$_5$. Our mass is computed with respect to pure AdS$_5$.

We verified that these quantities obey the first law of black hole thermodynamics $\mathrm{d}M=T\mathrm{d}S+3\mu \mathrm{d}Q$ at least to $0.01\%$.

\section{\label{sec:results}Results}

\subsection{Phase diagram of hairy AdS$_5\times S^5$ black holes}
\label{subsec:Phase} 

In this subsection we present a comprehensive picture of the phase diagram of the hairy black holes in the microcanonical ensemble and analyse its rich structure.

\begin{figure}[t]
\centering
  \begin{minipage}[t]{0.45\textwidth}
    \includegraphics[width=\textwidth]{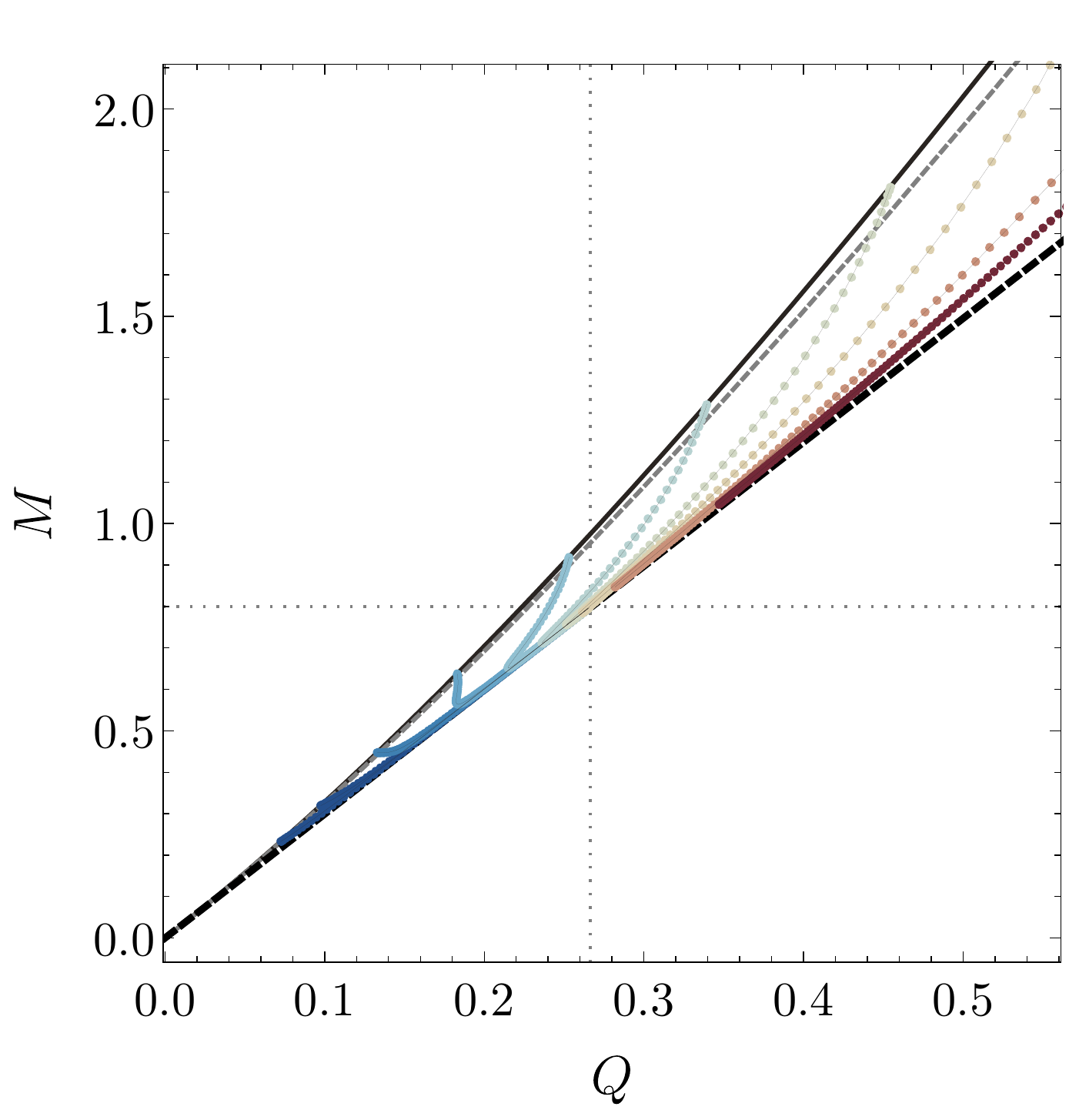}
  \end{minipage}
 \hfill \hfill
  \begin{minipage}[t]{0.53\textwidth}
    \includegraphics[width=\textwidth]{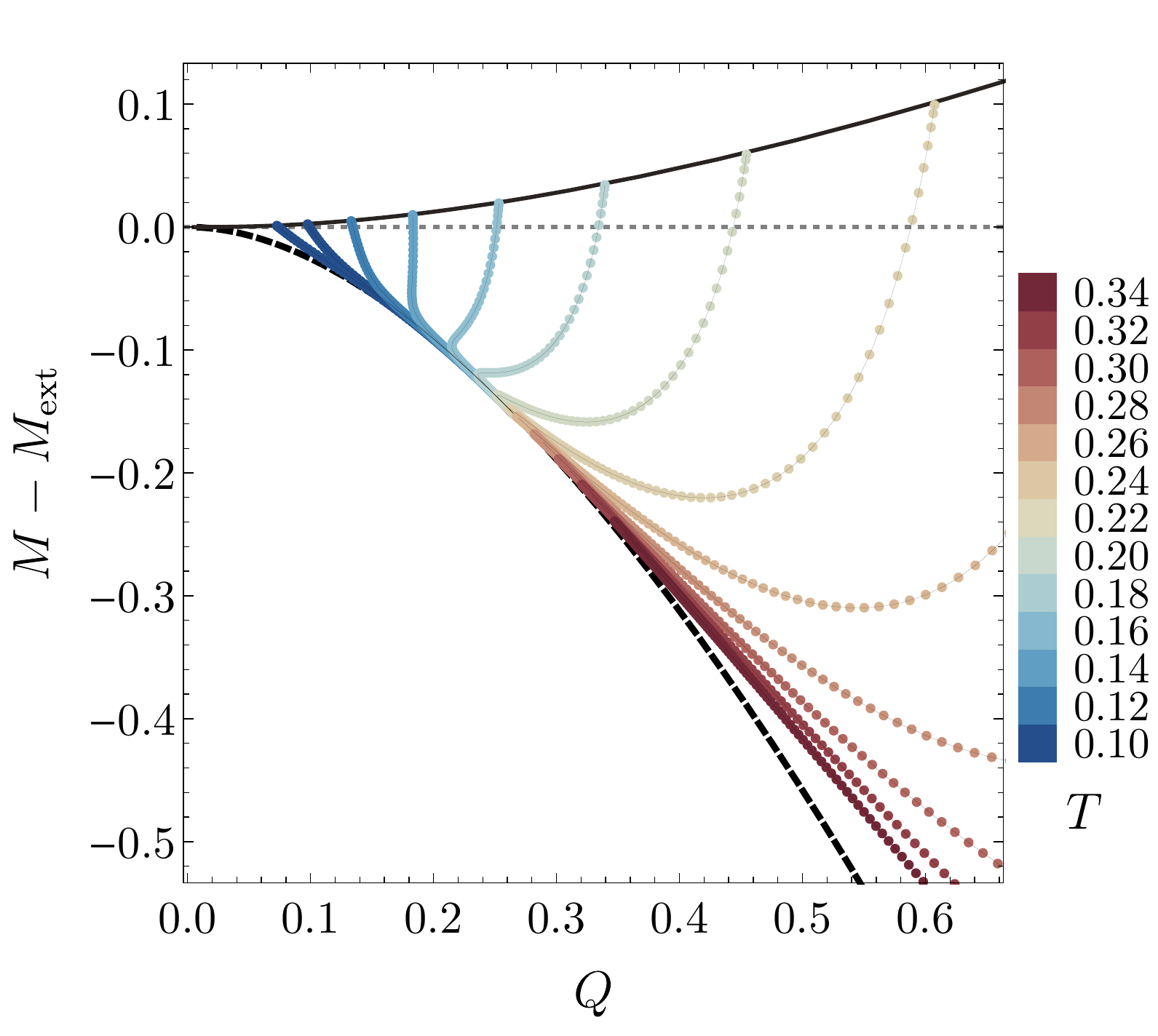} 
  \end{minipage}
  \caption{\textit{Left}: Phase diagram for the hairy black holes. The merger curve (solid black) indicates the onset of the superradiant instability. The line of extremal RNAdS solutions is shown as a dashed gray line. The BPS bound is given by $M_\mathrm{BPS}(Q)=3Q$ (dashed black). The gray dotted gridlines indicate the position of the special soliton with $\alpha = 2/3$. \\
  \textit{Right}: For clarity, we plot the mass difference $\Delta M=M-M_{\mathrm{ext}}$, where $M_{\mathrm{ext}}$ is the mass of an extremal RNAdS black hole with the same charge $Q$.}
  \label{fig:mq}
\end{figure}

The black holes with a non-zero scalar condensate first start to exist where the RNAdS black holes become superradiantly unstable. The RNAdS black holes are uniquely specified by the two parameters ($R$, $\mu$) and given the horizon radius we look for the value of the chemical potential $\mu$ at which zero-mode of the scalar field first appears. We generate this one parameter family of solutions separately by linearising the scalar equation~\eqref{eq:a:scalar} in the compact variable $y$ around the RNAdS black hole. Let $\delta q_5$ be an infinitesimal perturbation of $q_5$ defined in (\ref{eq:ansatz1gauge}). Following \cite{Dias:2010ma, Dias:2011tj} we numerically solve the resulting generalised eigenvalue problem
\begin{equation}
L(y)\delta q_5(y)=\mu^2\Lambda(y)\delta q_5(y) 
\end{equation}
with boundary conditions $\delta q_5^\prime(0)=0$ and $2\mu^2\delta q_5(1)-R^2\delta q_5^\prime(1)=0$ which follow from imposing regularity at the horizon and solving \eqref{eq:a:scalar} near the asymptotic infinity. The $L(y)$ and $\Lambda(y)$ are both second order differential operators independent of $\mu$. The chemical potential of the corresponding marginally stable RNAdS black hole appears as the generalised eigenvalue. The line of solutions representing the onset of the condensation is also obtained by solving the full non-linear equations of motion~\eqref{eq:eom} setting $q_5(0)=\epsilon_0$ to be small. These two methods to generate the merger line are found to be in very good agreement.

Our numerical results are presented in Fig.~\ref{fig:mq}. We find that the hairy black holes exist between the instability curve all the way down to the BPS bound and we verified it for a wide range of charges. Numerically we did not find any upper bound on the charge up to $Q\sim 100$ and from the structure of the phase diagram it would be natural to infer that the hairy black hole solutions exist between the merger line and the BPS bound for every charge. At the lower bound the hairy black holes join the solitonic solution in the phase diagram, in particular, in the limit $T\rightarrow 0$, hairy black holes approach the smooth soliton, just as predicted in \cite{Bhattacharyya:2010yg}.

\begin{figure}[t]  
\centering
  \begin{minipage}[t]{0.42\textwidth}
    \includegraphics[width=\textwidth]{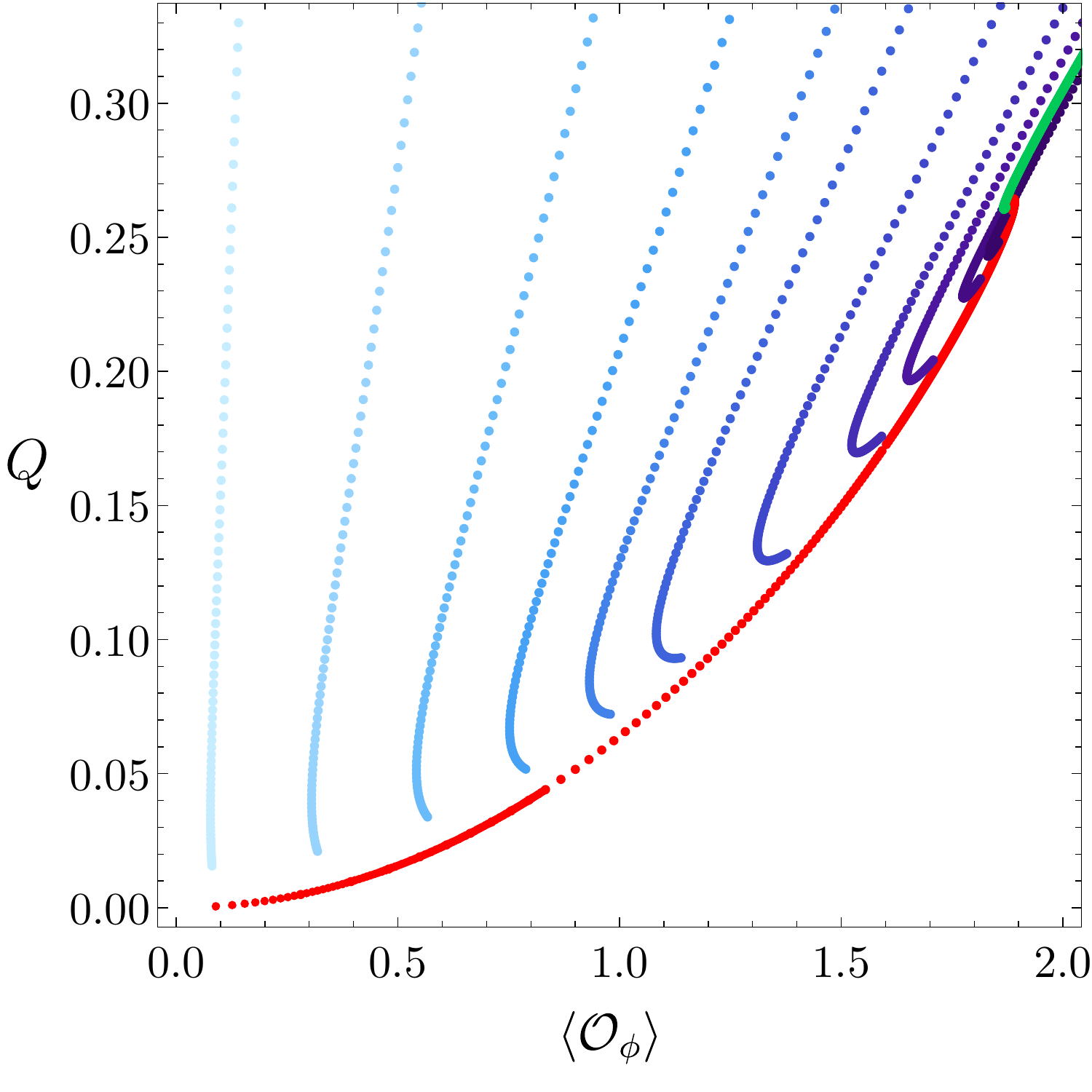}
  \end{minipage}
\hfill
  \begin{minipage}[t]{0.5\textwidth}
    \includegraphics[width=\textwidth]{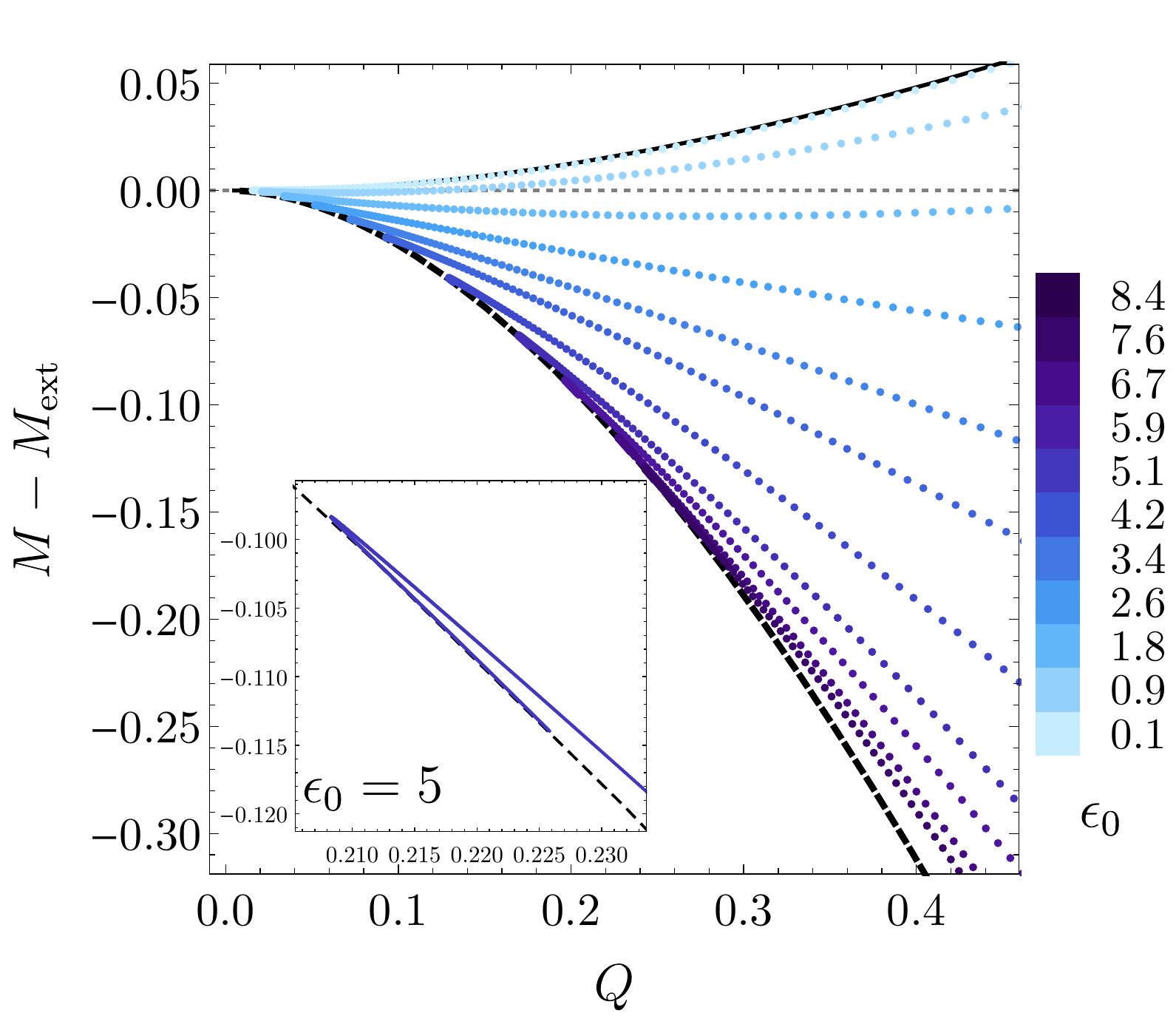}
  \end{minipage} 
\caption{\textit{Left}: The hairy black hole charge $Q$ versus $\langle\mathcal{O}_\phi\rangle$ for constant central scalar field density $\epsilon_0$ curves. Red line is the smooth soliton and the green line is the singular soliton. \textit{Right}: Mass difference versus charge $Q$. The constant parameter $\epsilon_0$ curves extend down to $T=0.055$. The inset is a zoomed in plot around $Q=Q_c$ for some value of $\epsilon_0$.}
    \label{fig:consteps}
\end{figure}

In more detail, in Fig.~\ref{fig:consteps} we plot the charge $Q$ as a function of $\langle\mathcal{O}_\phi\rangle$ for constant values of $\epsilon_0$. In order to parametrise each of these constant $\epsilon_0$ curves we dial the temperature $T$. As we lower the temperature, we see that hairy solutions join smoothly to the smooth soliton curve. Furthermore, the higher value of $\epsilon_0$ we choose, the closer the hairy solutions get to $Q=Q_c$. In particular, as $\epsilon_0\to+\infty$ we see that the hairy black hole solution inherits the spiralling behaviour of the smooth solitonic branch (see right panel of Fig.~\ref{fig:consteps} where we can see two arms of the spiral). Note that exactly for $\epsilon=\langle O_\phi \rangle = 1.8710$ we expect even the hairy black hole to have infinite non-uniqueness as we approach $T\to0$ from above.

\begin{figure}[t]
\centering
  \begin{minipage}[t]{0.44\textwidth}
    \includegraphics[width=\textwidth]{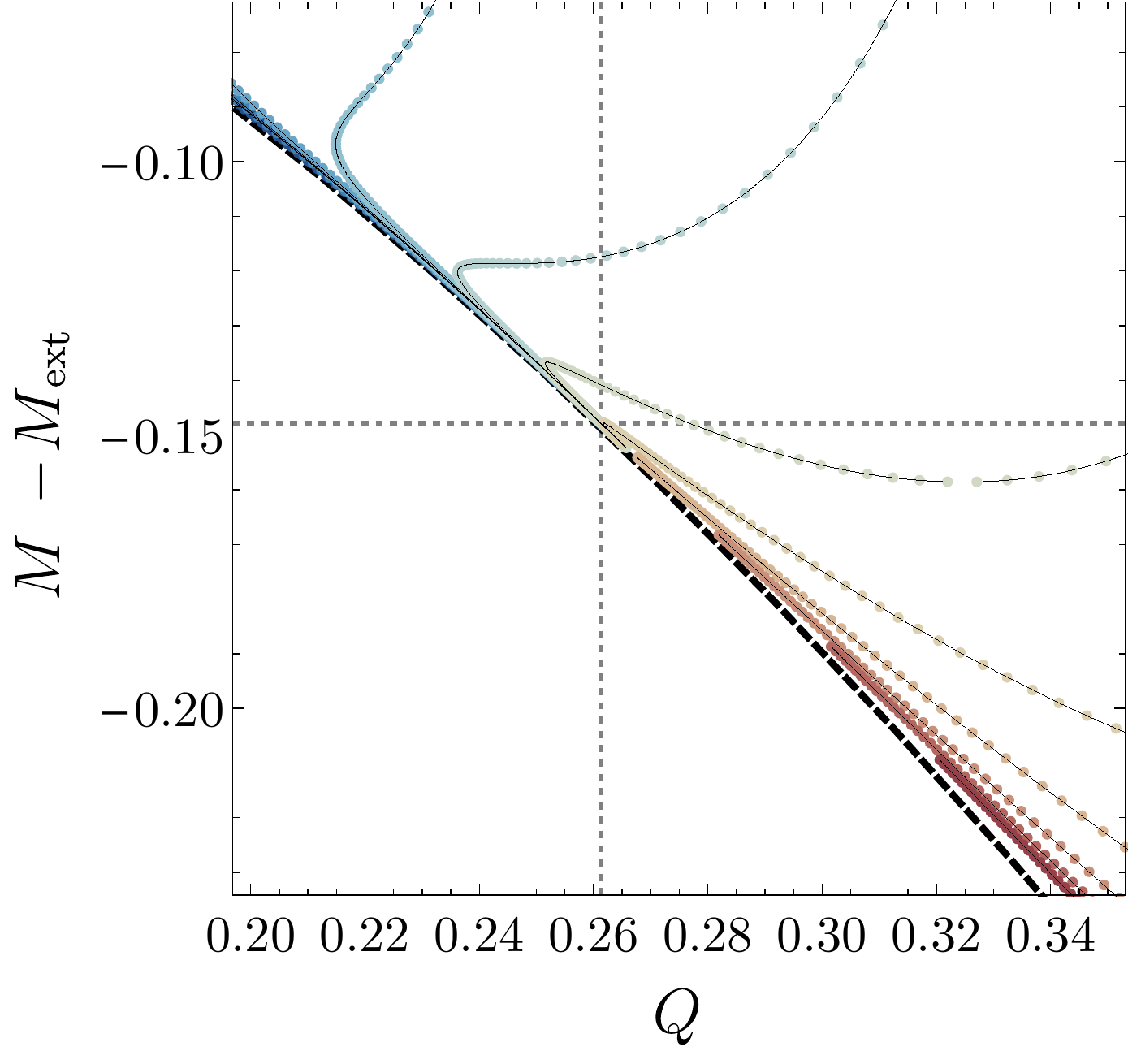}
  \end{minipage} 
 \hfill
  \begin{minipage}[t]{0.53\textwidth}
    \includegraphics[width=\textwidth]{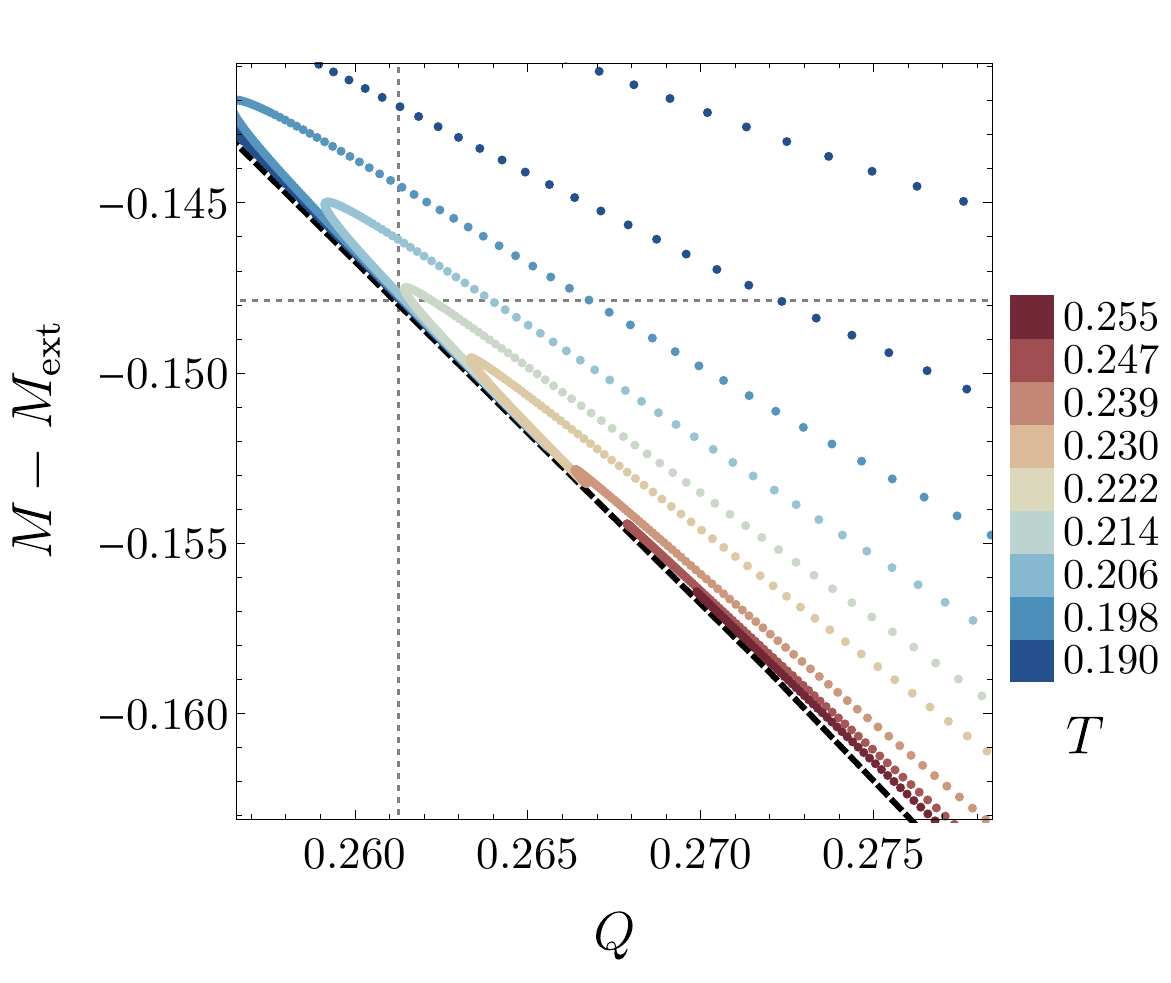}
  \end{minipage}
    \caption{\textit{Left}: Zooming in around $Q=Q_c$, and observing the transition between $T<T_2$ and $T>T_2$. The color legend is the same as in Fig.~\ref{fig:mq}. \textit{Right}: An even closer look for $Q$ near $Q_c = 0.261$. The hairy black hole isotherms terminate at charges above the special singular soliton.}
    \label{fig:zoom}
\end{figure}

The behaviour of the isothermal curves changes as a function of the temperature. In particular, if we fix a temperature in the interval $T_1<T<T_2$ while increasing $\epsilon_0$, with $T_1 = 0.139^{+0.002}_{-0.002}$ and $T_2 = 0.23^{+0.01}_{0}$, we find two solutions for the same value of the charge $Q$ (corresponding to two different values of $\epsilon_0$). This can be seen for instance on the left panel of Fig.~\ref{fig:zoom}. We shall shortly see that this feature will give an intricate phase diagram in the canonical ensemble, where the temperature and charge are held fixed.

\begin{figure}[t]
\centering
  \begin{minipage}[t]{0.47\textwidth}
    \includegraphics[width=\textwidth]{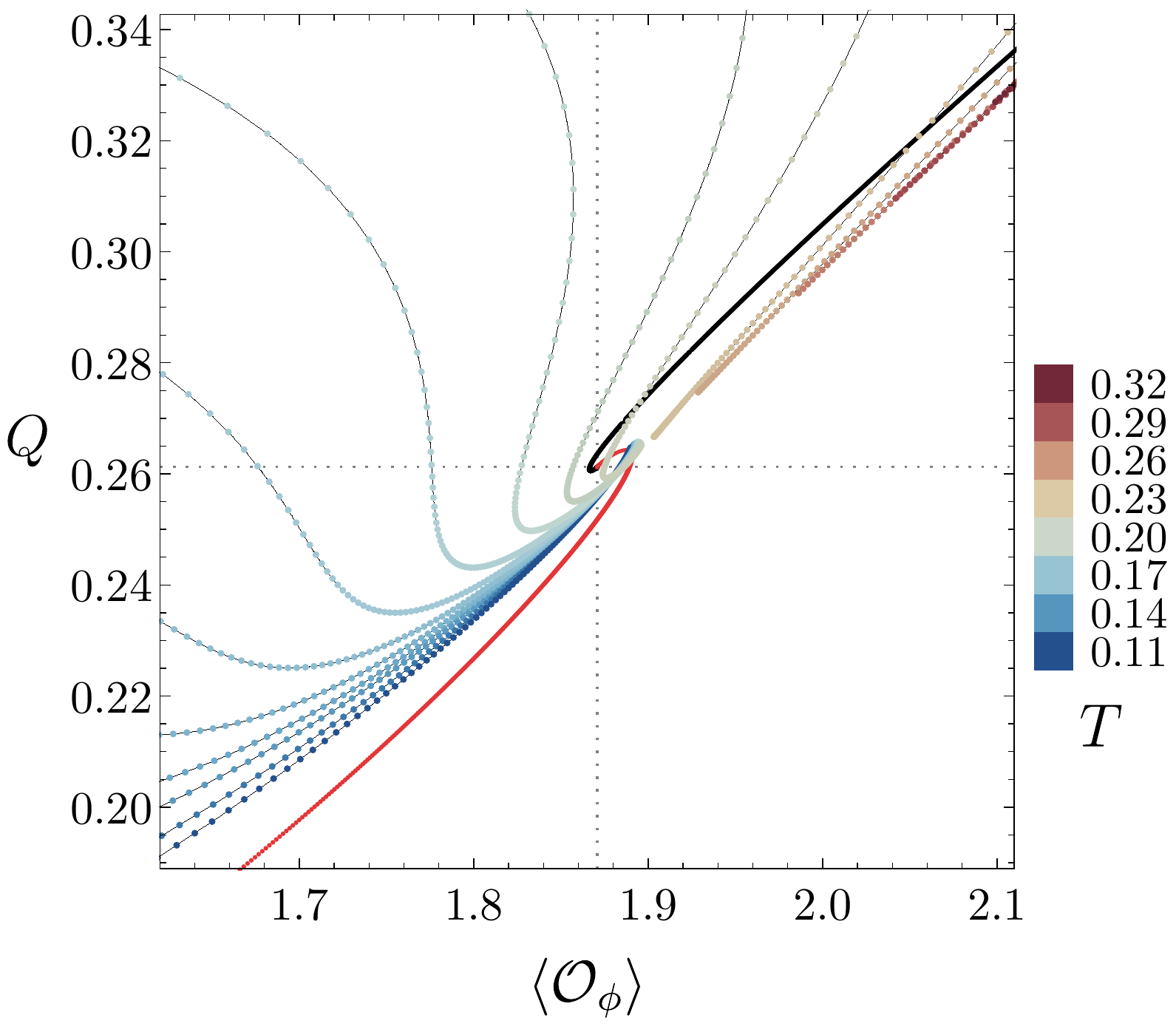}
  \end{minipage}
\hspace{+1em}
  \begin{minipage}[t]{0.42\textwidth}
    \includegraphics[width=\textwidth]{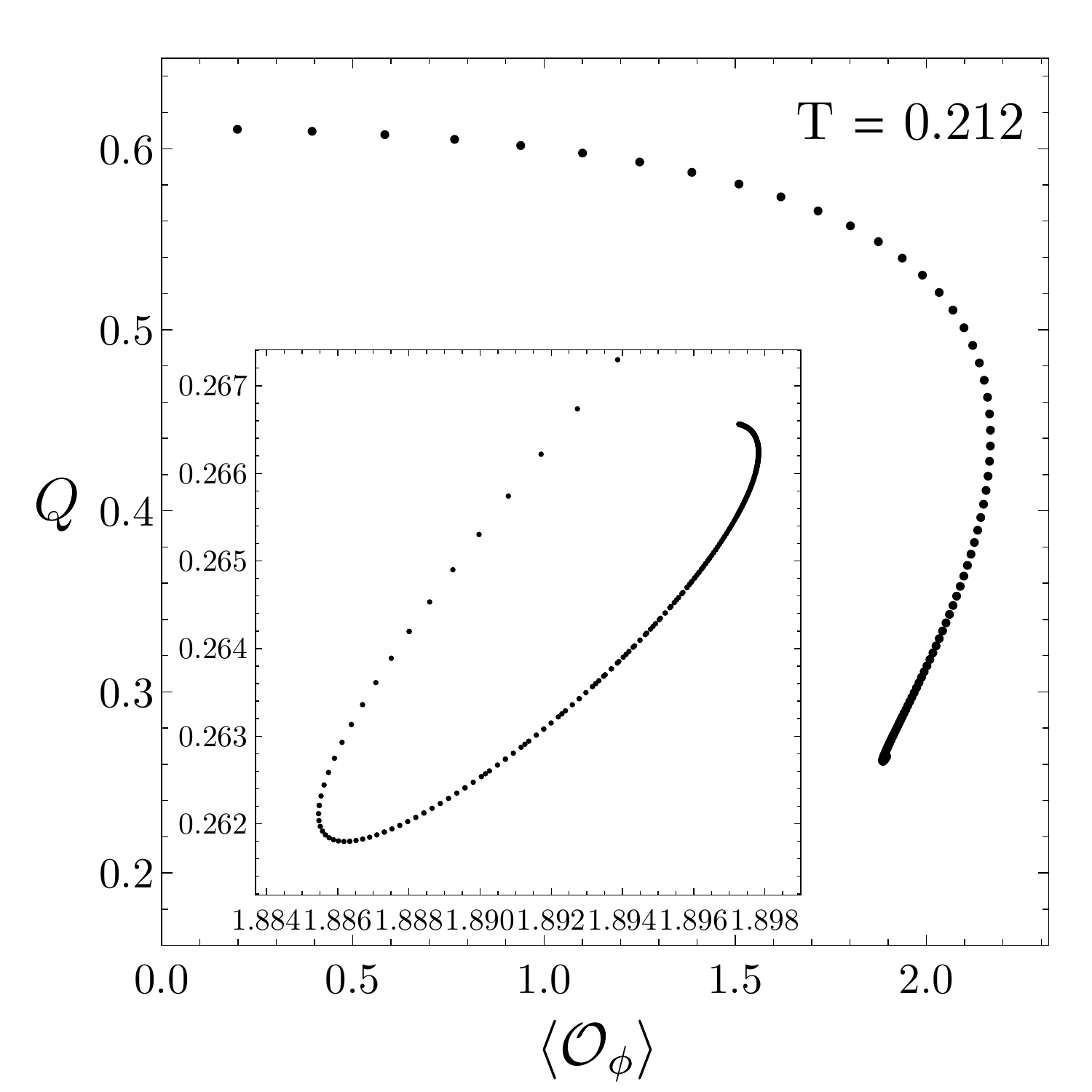}
  \end{minipage} 
\caption{\textit{Left}: Charge versus the vacuum expectation value of the operator dual to the scalar field for constant temperature hairy black hole solutions. The black and red data points are singular and smooth solitons respectively. Dotted gridlines show the point where these two merge. \textit{Right}: The charge of the hairy solutions as we approach the singular soliton with $\alpha=1$, exhibits damped oscillations. This data was collected with $n=1000$ grid points.}
    \label{fig:spiral}
\end{figure}

One can finally ask what is the fate of the isothermal curves as we increase $\epsilon_0$. According to what we described above, these cannot be connected to the smooth soliton (except for the special isothermal with $T=0$). Indeed, we find numerical evidence that they connect to the singular soliton with $\alpha=1$, see for instance the right panel of Fig.~\ref{fig:zoom} where we see constant temperature curves joining the BPS bound at $M_\star=3Q_\star>Q_c$, with the limiting $Q_\star$ increasing further away from $Q_c$ as we increase the temperature. This behaviour can also be seen on the left panel of Fig.~\ref{fig:spiral}. Finally, we note that as the hairy black hole isothermals approach the singular soliton, we find evidence for spiralling behaviour, which is depicted on the right panel of Fig.~\ref{fig:spiral}.

\begin{figure}[]
\centering
  \begin{minipage}[t]{0.48\textwidth} 
    \includegraphics[width=\textwidth]{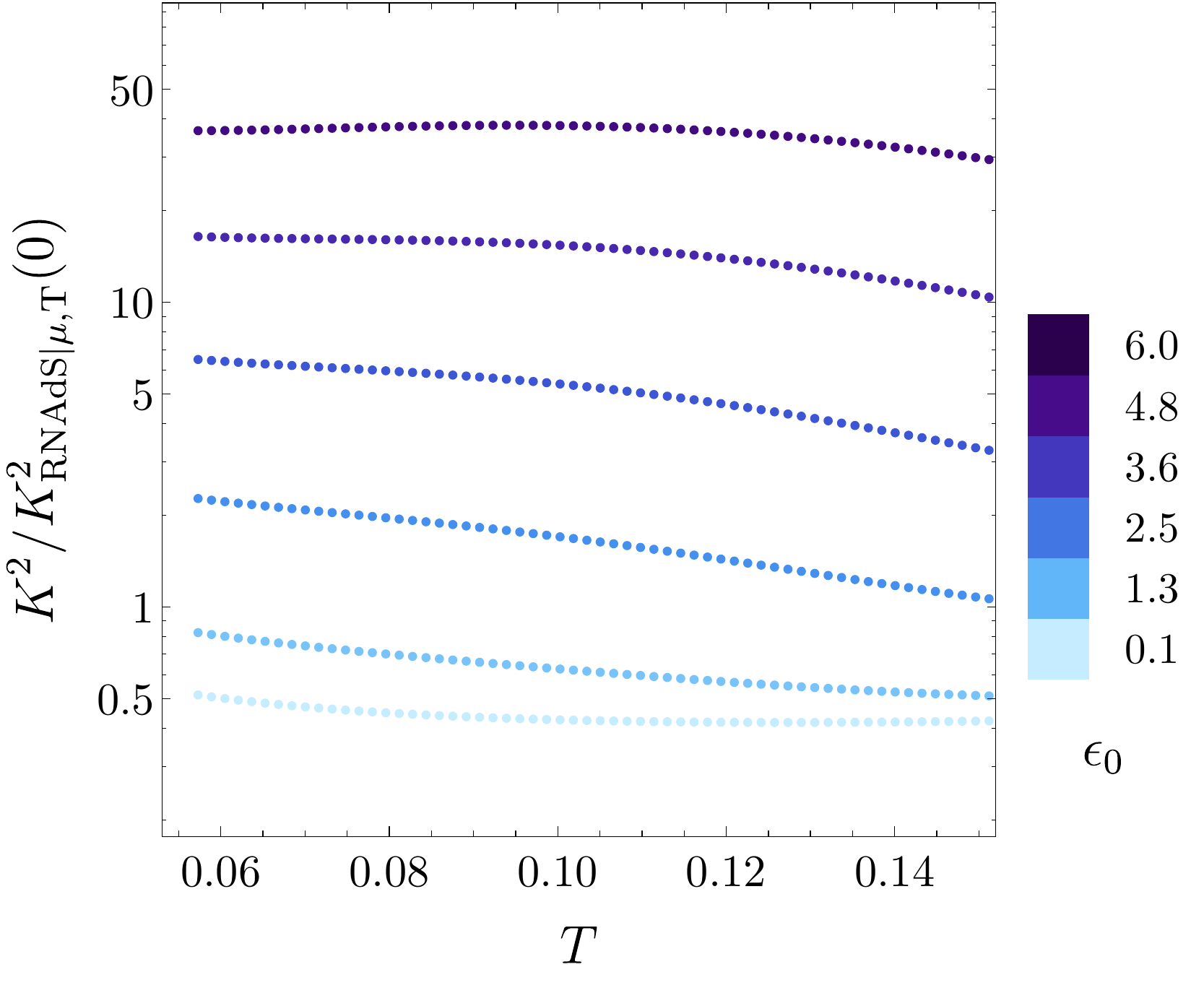}
  \end{minipage} 
  \hfill
    \begin{minipage}[t]{0.48\textwidth}  
    \includegraphics[width=\textwidth]{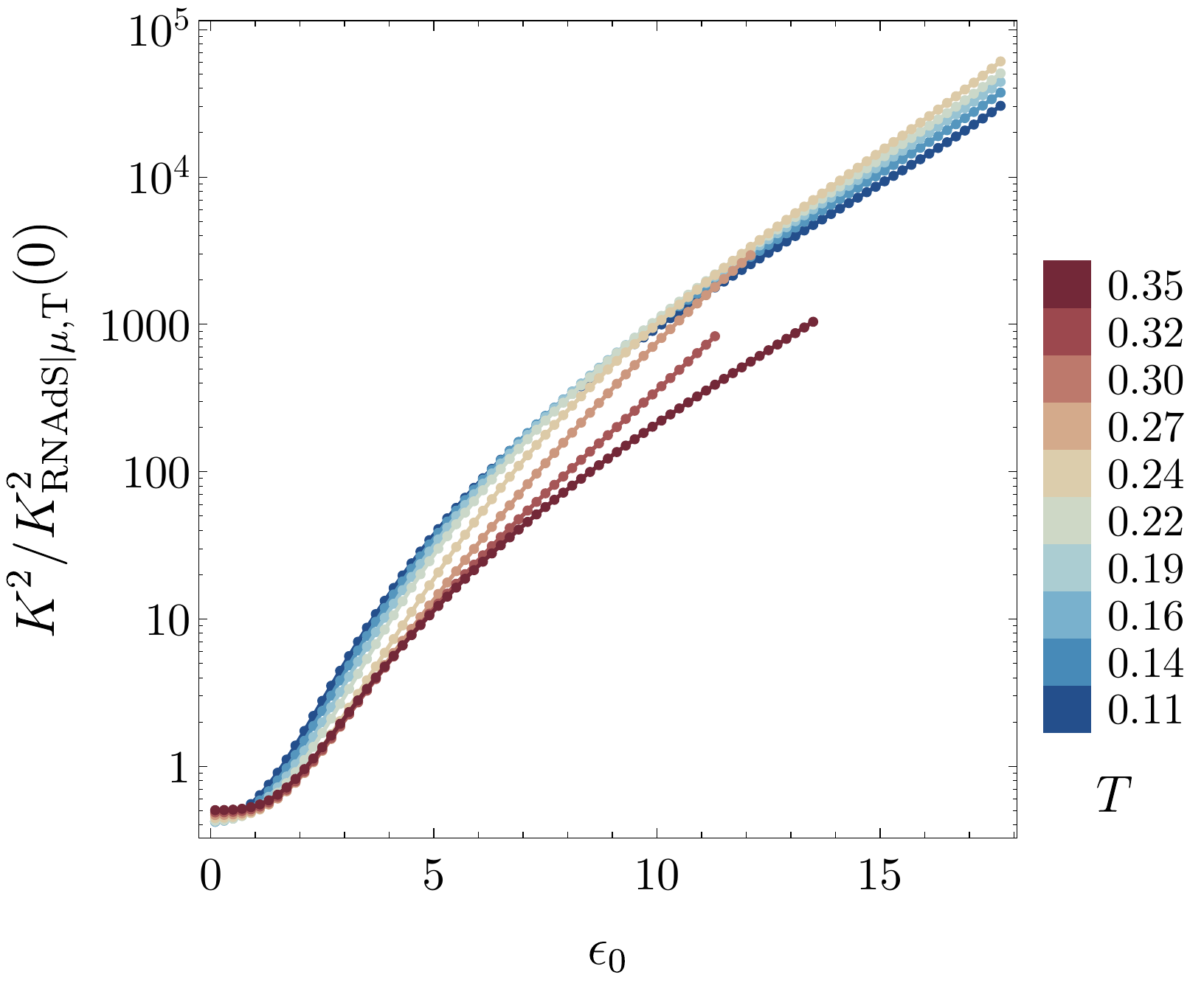}
  \end{minipage} 	
\caption{
 \textit{Left}: Curvature invariants at the origin for a range of temperatures for constant $\epsilon_0$. The Kretchmann scalar remains finite. For lower values of $\epsilon_0$ it takes longer for the hairy black holes to approach the BPS bound, hence for larger values of the parameter the curves flatten out quicker.
 \textit{Right}: Kretchmann invariant $K^2$ for the range of temperatures scaled by the $K^2$ of the corresponding RNAdS black hole in the grand-canonical ensemble. As $\epsilon_0$ increases the invariant increases without a bound.}
    \label{fig:invariant}
\end{figure}

In order to support the claim that $T\rightarrow 0$ hairy black holes do not tend to some configuration possessing irregular geometry we compute the Kretschmann invariant $K^2=R_{abcd}R^{abcd}$ following \cite{Dias:2011tj}. Because for RNAdS $K^2\sim 1/R^4$ when $T\rightarrow 0$, we normalise the $K^2$ by that of the corresponding RNAdS black hole with the same chemical potential and temperature. In Fig.~\ref{fig:invariant} (left) we show that the normalised curvature invariant remains bounded as we approach the smooth soliton.  

On the other hand, keeping the temperature fixed and increasing $\epsilon_0$ the normalised Kretschmann invariant appears to blow up (see Fig.~\ref{fig:invariant}, right) as we approach the BPS bound. For solutions with $T>T_2$ we found the metric \emph{ansatz} \eqref{eq:metric1} to be more numerically stable.

As we increase $\epsilon_0$ the hairy black hole isotherms are approaching the BPS bound. The chemical potential $\mu\rightarrow 1$ and the entropy $S\rightarrow 0$ (see Fig.~\ref{fig:entropies}, Appendix~\ref{sec:therm}). This together with the fact that the Kretschmann invariant blows up as $\epsilon_0\to +\infty$, even when normalized by the Reissner-Nortstr\"om solution, suggests that the isothermals will merge with the $\alpha=1$ soliton, for any value of $T$.
 
\subsection{The planar limit}
\label{subsec:planar}
In this subsection we consider the planar horizon limit of our global AdS$_5$ solutions. The resulting black brane solutions were first studied in great detail in \cite{Aprile:2011uq}. Our numerical approach is similar to the one we used for the spherical black holes, so here we just quote the final results. In the large charge limit the singular soliton branch admits an exact analytical solution from which the planar limit solution can be recovered \cite{Bhattacharyya:2010yg}
\begin{equation}
\label{eq:exact}
\mathrm{d}s^2=-r^2\mathrm{d}t^2+\frac{r^2\mathrm{d}r^2}{b^2+r^4}+r^2\mathrm{d}x^2,\quad\phi(r)=\frac{2b}{r^2},\quad A(r)=0.
\end{equation}
Note that the choice of the constant $b$ amounts to a coordinate transformation, therefore, this is a single asymptotically Poincar\`e patch solution. The planar solution exhibits explicit conformal invariance, since the field theory is suppose to live on Minkowski spacetime. Thus, in order to have a well defined planar limit, we always look at conformal invariant ratios, which should have a smooth limit as the black holes become infinitely large. For instance, to measure temperature we introduce  $\tilde{T}\equiv\sqrt{\varepsilon_0} T$. The planar hairy solutions are thus a one parameter family of solutions, with the singular soliton solution (\ref{eq:exact}) being a point. We choose to parametrise the hairy branes by $\tilde{T}$.

\begin{figure}[t]
\centering
\includegraphics[width=1\textwidth]{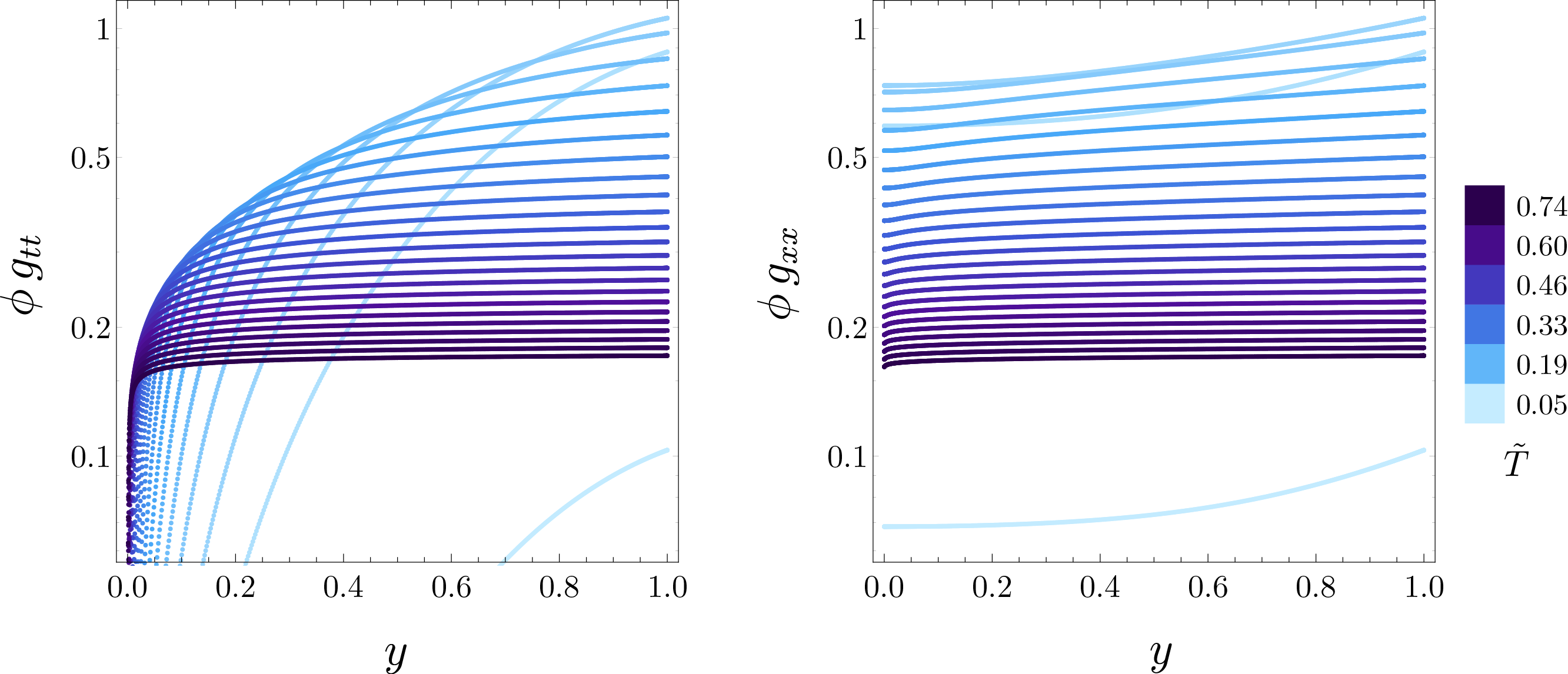}
\caption{Gauge invariant quantities $\phi g_{xx}$ and $\phi g_{rr}$ for the hairy planar solutions, versus the compact coordinate $y$. According to the exact solution, both these curves should approach the same constant value when $\tilde{T}\to+\infty$.}
\label{fig:functions}
\end{figure}

We have constructed planar hairy black holes, \emph{i.e.} hairy black branes, and checked that our spherical hairy black holes do approach the hairy branes in the limit $S\to+\infty$, \emph{i.e.} hairy black holes become infinitely large. Furthermore, the singular soliton solution (\ref{eq:exact}) is the zero $M/T^4$ and $Q/T^3$ limit of the hairy black branes, see Fig.~\ref{fig:pqm}, which is reached as we take $\tilde{T}$ to be large. In order to test this, we have plotted the following gauge invariant quantities $g_{tt}\phi$ and $g_{xx}\phi$ and checked that they approach the same value at large $\tilde{T}$, see Fig.~\ref{fig:functions}. This is what is predicted by the exact soliton solution (\ref{eq:exact}). 

\subsection{Thermodynamics} 
\label{subsec:therm}

In this section we analyse phase diagrams arising in different thermodynamic ensembles. The planar limit of our results match the results of \cite{Aprile:2011uq}, which we reproduced using our own code. For completeness we present in Fig.~\ref{fig:planartherm} of the Appendix \ref{sec:therm}, a complete analysis of the several ensembles in the planar limit. In particular, we find that the hairy black branes are only dominant in the microcanonical ensemble, but never in the canonical or grand-canonical ensembles. 

\subsubsection{Grand-canonical ensemble}

\begin{figure}[t]
  \begin{minipage}[t]{0.45\textwidth}
    \includegraphics[width=\textwidth]{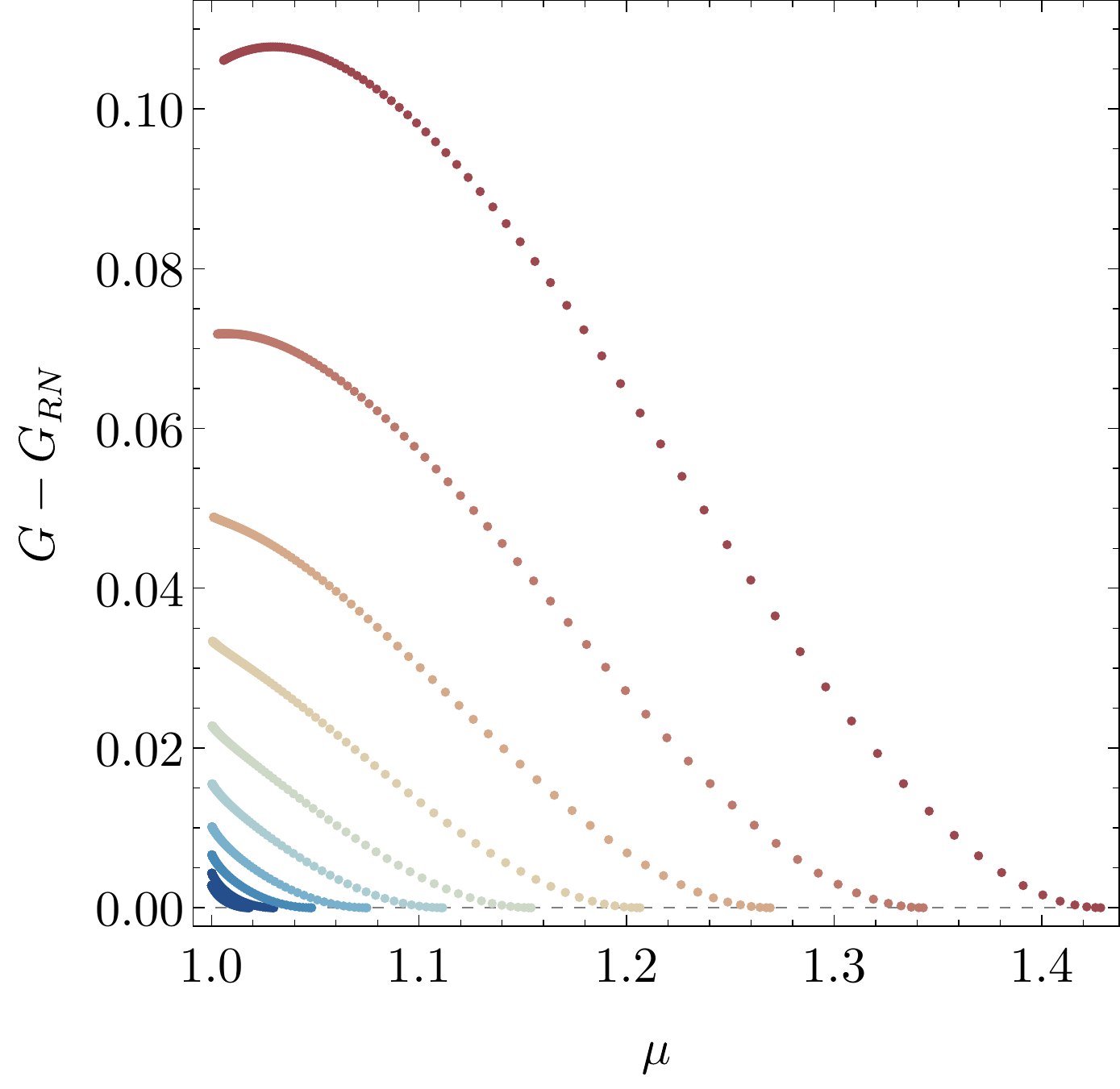}
  \end{minipage}
  \hfill
  \begin{minipage}[t]{0.53\textwidth} 
    \includegraphics[width=\textwidth]{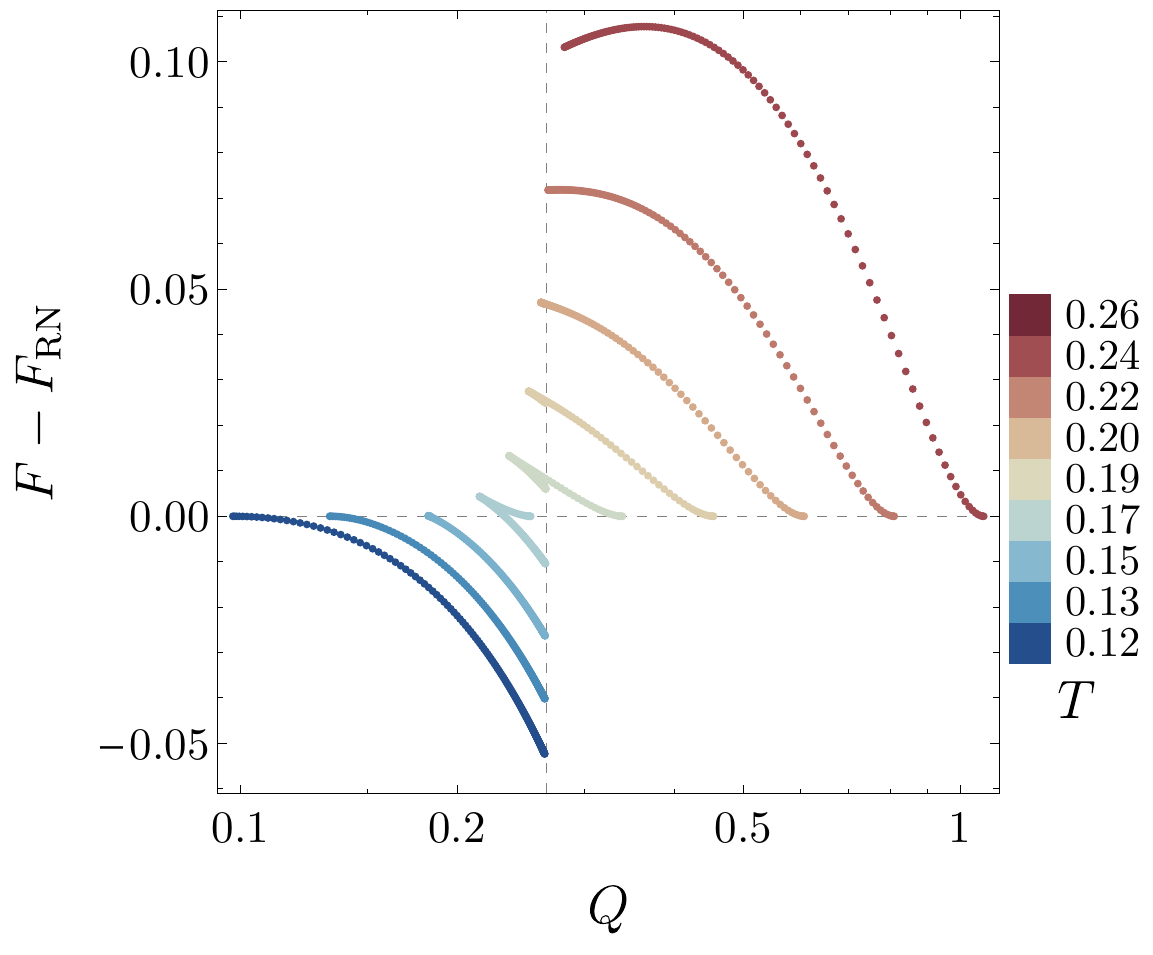} 
  \end{minipage}
      \caption{\textit{Left}: The difference between the Gibbs free energies of hairy and RNAdS solutions with the same chemical potential and temperature. \textit{Right}: Difference of the Helmoltz free energies of the Reissner-Nortstr\"om and the hairy solution with the same temperature and charge.}
   \label{fig:gibbs}
\end{figure}

In the grand-canonical ensemble the system is in equilibrium with a thermodynamic reservoir with a temperature $T$ and chemical potential $\mu$, but is allowed to exchange energy and electric charge. The preferred phase of such a system minimises the Gibbs free energy $G=M-TS-3\mu Q$. The results are presented in the left panel of Fig.~\ref{fig:gibbs} as a difference between hairy black holes and RNAdS potentials (absolute quantities for a few regions in moduli space are shown in the Appendix~\ref{sec:therm}, Fig.~\ref{fig:gibbsfull}). We find that in the grand-canonical ensemble RNAdS black holes have lower Gibbs free energy than the hairy black holes with the same chemical potential $\mu$ and temperature $T$. Note that our hairy solutions all have $G<0$, so $G>G_{RN}$ and that the RNAdS black holes and the hairy black holes phases are identical at the merger points. For RNAdS, $G=\frac{1}{4}R^2(1-R^2-\mu^2)$ and as shown in the Fig.~\ref{fig:gibbsfull} (Appendix~\ref{sec:therm}) the Gibbs free energy for the hairy black holes is always negative and approaching $0$ as we increase $\epsilon_0$. Note that it would only be exactly zero if $\mu$ could reach $1$, but that can only happen at infinite $\epsilon_0$. Finally, so far we have only considered the transition between the hairy black holes and RNAdS black holes. However, we note that the RNAdS black holes can themselves become subdominant with respect to AdS \cite{Hawking:1982dh}. As all our energies are measured with respect to pure AdS, so that the energy of AdS simply corresponds to $M=0$ and therefore zero thermodynamic potentials, black holes with negative free energy are thermally favoured over pure AdS. The small RNAdS branch has $\mu\leq 1$ and thus these black holes never compete with the hairy solutions.

We have also studied local thermodynamic stability of the hairy black holes in the grand-canonical ensemble. We find that the specific heat at constant chemical potential is always positive, but the isothermal capacitance, defined as:
\begin{equation}
C_T = \left(\frac{\partial Q}{\partial \mu}\right)_T\,,\nonumber
\end{equation}
exhibits an interesting behaviour. For $T>T_2$ it is always positive and for $T<T_1$ we find $C_T<0$. In the interval $T_1<T<T_2$, each isothermal has two solutions at fixed electric charge. The most energetic of these solutions has $C_T>0$, whereas the least energetic has $C_T<0$.

\subsubsection{Canonical ensemble}

\begin{figure}[!t]
\centering
  \begin{minipage}[t]{0.44\textwidth}
    \includegraphics[width=\textwidth]{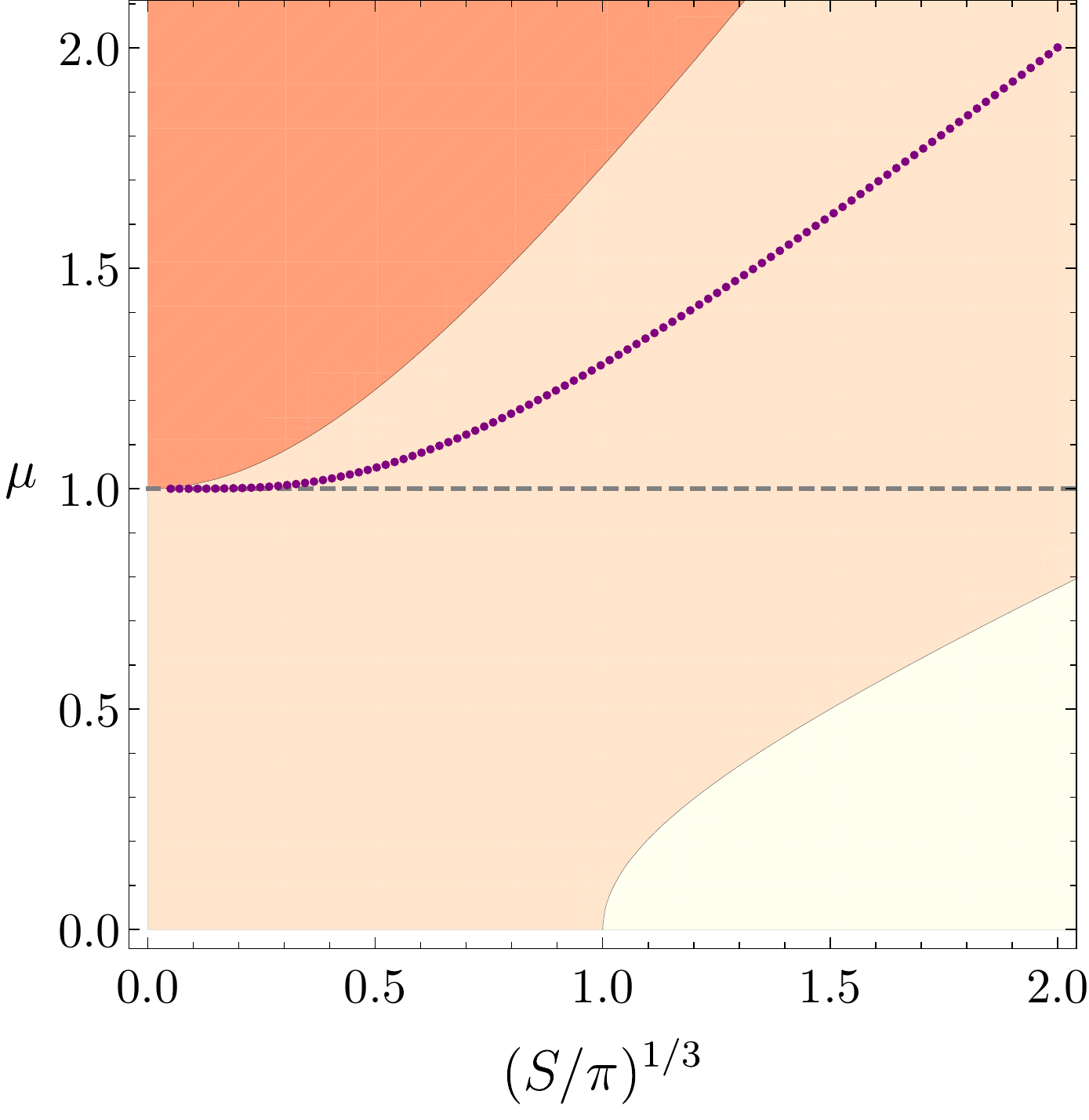}
  \end{minipage}
  \hfill
  \begin{minipage}[t]{0.44\textwidth}
    \includegraphics[width=\textwidth]{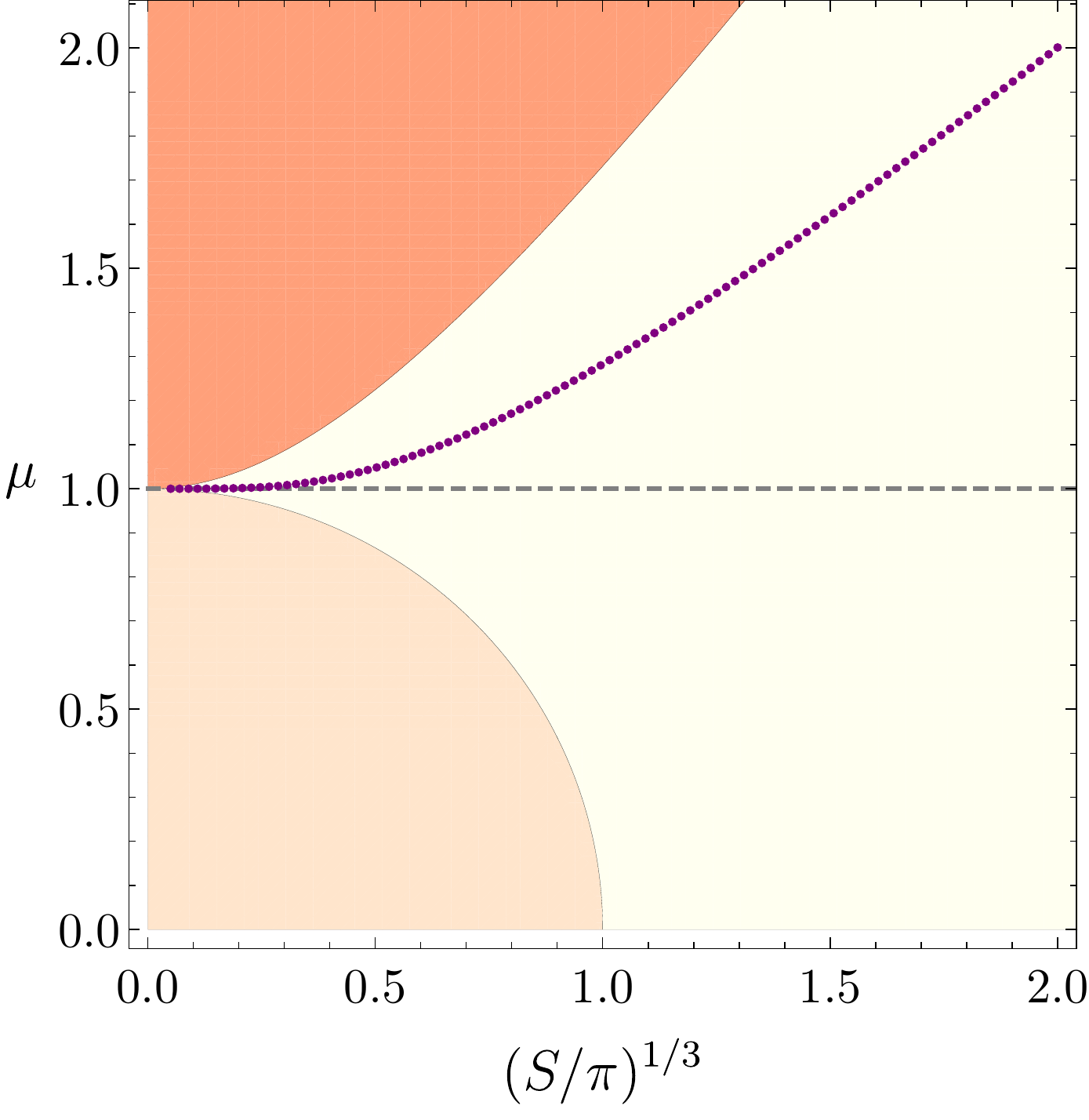}
  \end{minipage} 
      \caption{\label{fig:parameters}\textit{Left}: Canonical ensemble: RNAdS black holes exist below the extremality curve (the curve separating orange (top left) and light-orange (middle) regions) and the hairy black holes exist above the merger curve (purple data points) and for $\mu>1$. RNAdS dominate over pure AdS only in the yellow region (bottom right). The hairy black holes have a higher free energy than thermal AdS and thus are not the preferred phase in the ensemble. \textit{Right}: Grand-canonical ensemble: when both solutions coexist (again above the merger curve), the hairy black holes have a higher free energy than the corresponding RNAdS with the same $\mu$ and $T$. The yellow region (middle) shows the parameter space in which the RNAdS dominates over thermal AdS. The orange region (top left) is the extremal RNAdS and light-orange region (bottom left) is the sector in which pure AdS is preferred over the RNAdS black holes.} 
\end{figure}

In the canonical ensemble we restrict exchanges with the reservoir such that $\delta Q=0$, but $\delta M\neq0$, while keeping the temperature constant. The dominant phase minimises the Helmholtz free energy $F=M-TS$. The results are presented in the right panel of Fig.~\ref{fig:gibbs}. We see an interesting interplay between the RNAdS and the hairy solutions  which shows a phase transition in the constant temperature family of hairy solutions, occurring at $T_1$ and ending at $T_2$. The higher $\epsilon_0$ branch has lower $F$ than the corresponding RNAdS black hole, see Fig.~\ref{fig:helmfull} in Appendix~\ref{sec:therm}. For $T>T_2$ RNAdS has lower free energy than the hairy black hole. Note however that in the region where the hairy solutions dominate over the RNAdS black hole, $F$ is positive indicating that thermal AdS is the dominant phase in this region of moduli space. We have also studied the local thermodynamic stability of the hairy solutions in the region where they dominate over the corresponding RNAdS black holes. Local thermodynamic stability in the canonical ensemble is controlled by the sign of the specific heat at constant charge, which turns out to be positive for this range of $T$ and $Q$. We summarise our results for these two ensembles in Fig.~\ref{fig:parameters}.

\begin{figure}[!t] 
\centering
  \begin{minipage}[t]{0.44\textwidth}
    \includegraphics[width=\textwidth]{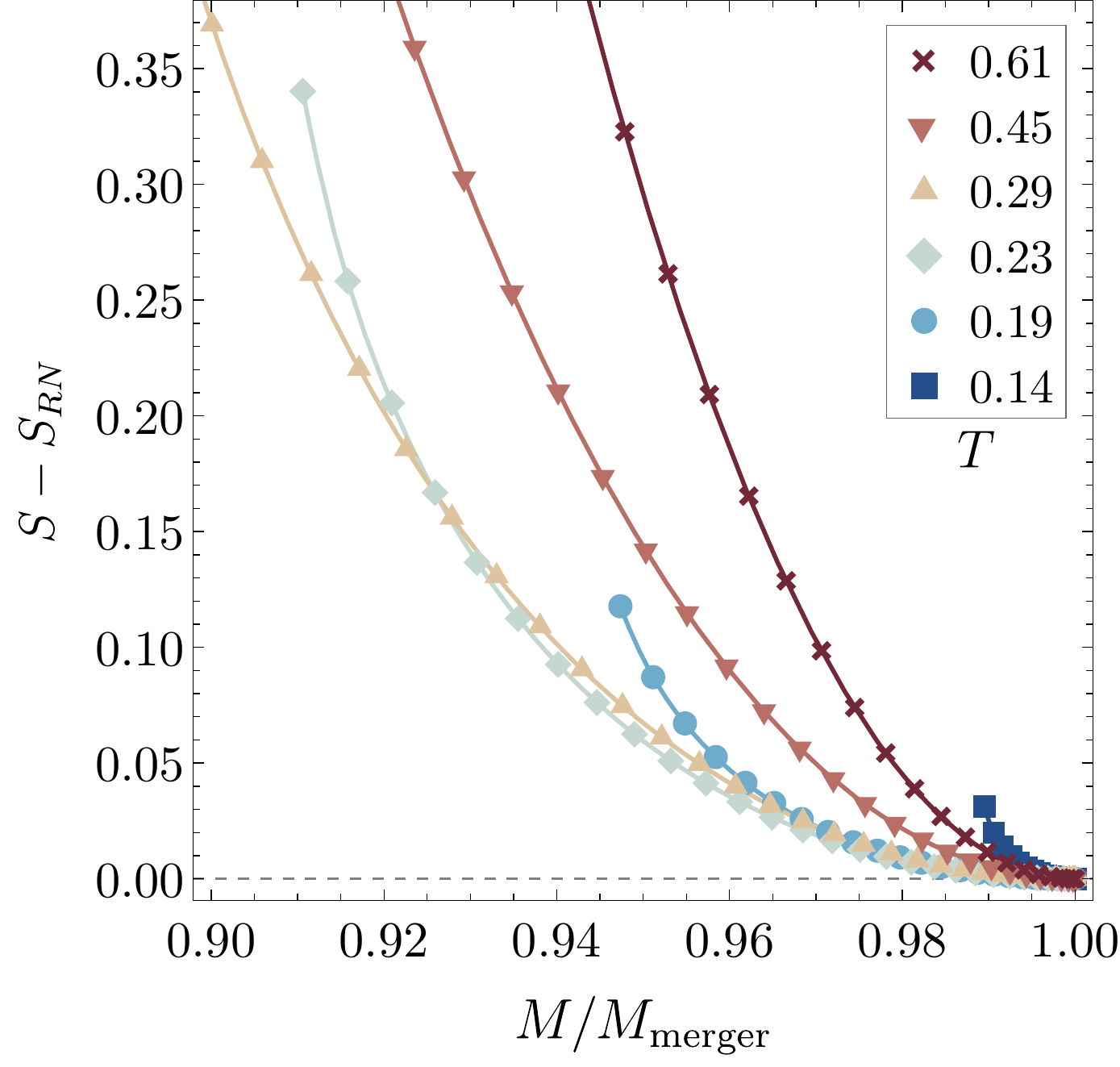}
  \end{minipage}
      \caption{The entropy difference $S-S_{\mathrm{RN}}$ of the hairy black holes and the corresponding RNAdS with the same values of $Q$ and $M$, as a function of the scaled mass $M/M_{\mathrm{merger}}$ plotted for a range of temperatures. Here $M_{\mathrm{merger}}$ is the mass at the onset of the superradiant instability, where by definition $S-S_{\mathrm{RN}}=0$.}
    \label{fig:entropy}
  \end{figure}
 
\begin{figure}[t]
\centering
  \begin{minipage}[t]{0.36\textwidth}
    \includegraphics[width=\textwidth]{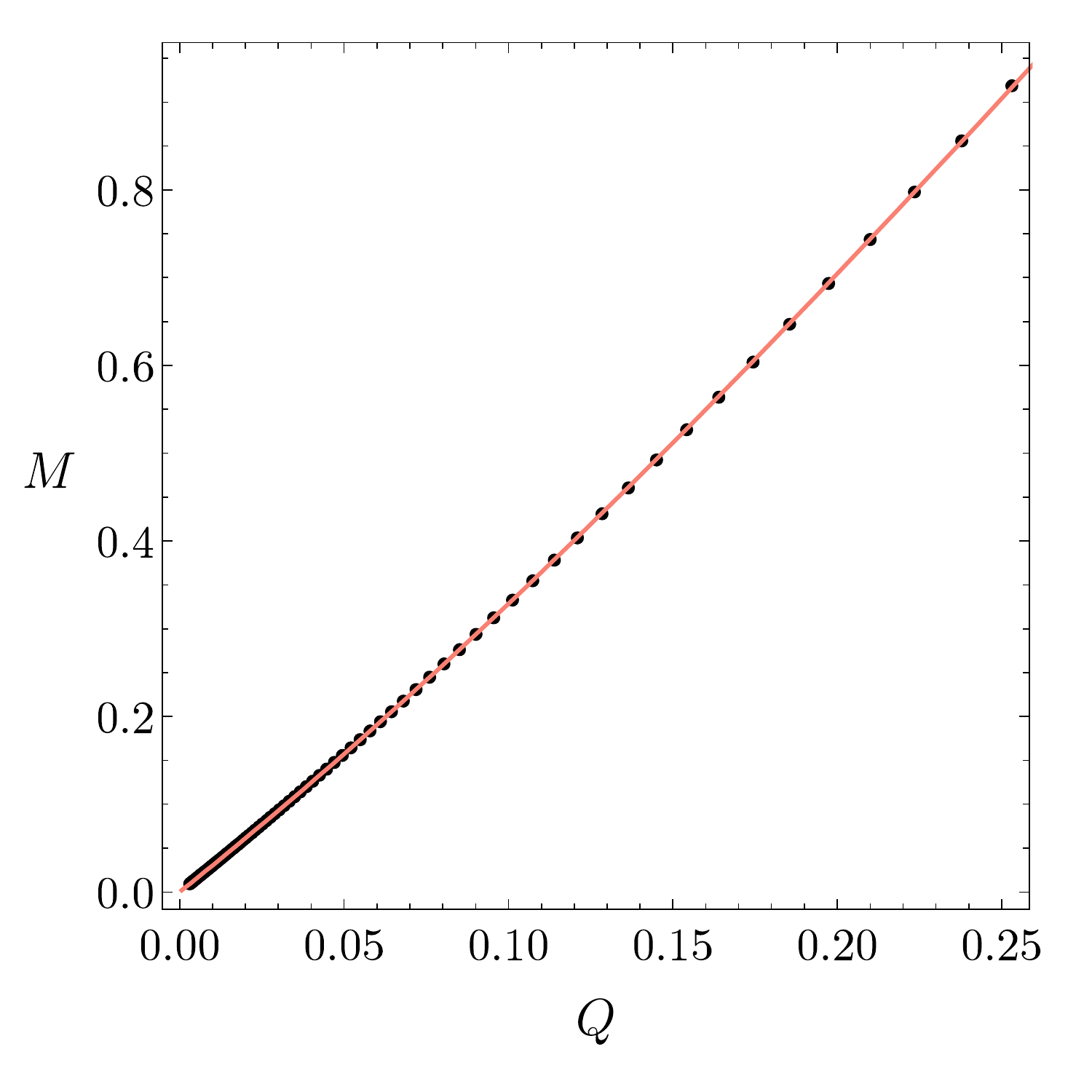}
  \end{minipage}
 \hspace{+2.8em}
    \begin{minipage}[t]{0.36\textwidth}
    \includegraphics[width=\textwidth]{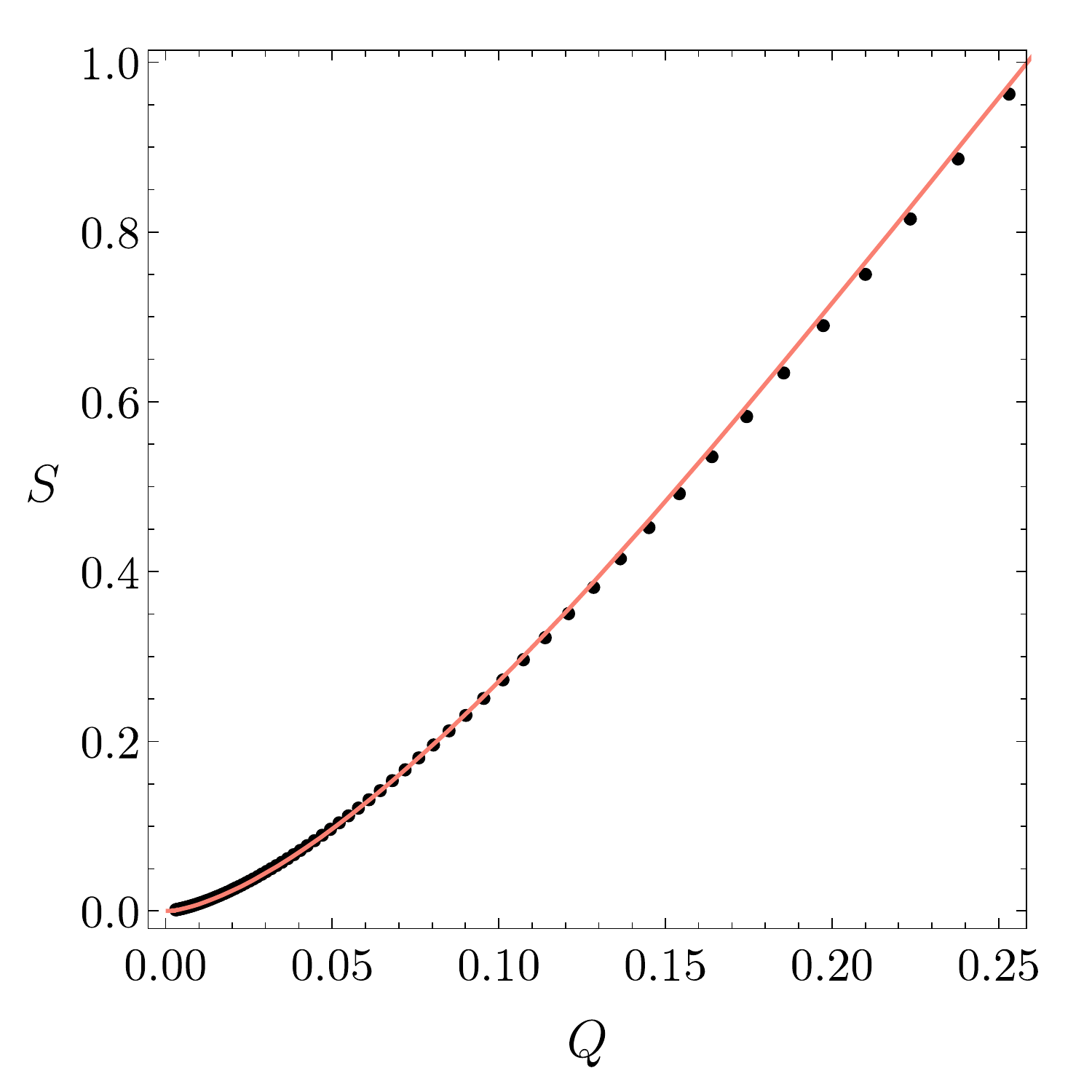} 
  \end{minipage}
  \vfill
    \begin{minipage}[t]{0.36\textwidth}
    \includegraphics[width=\textwidth]{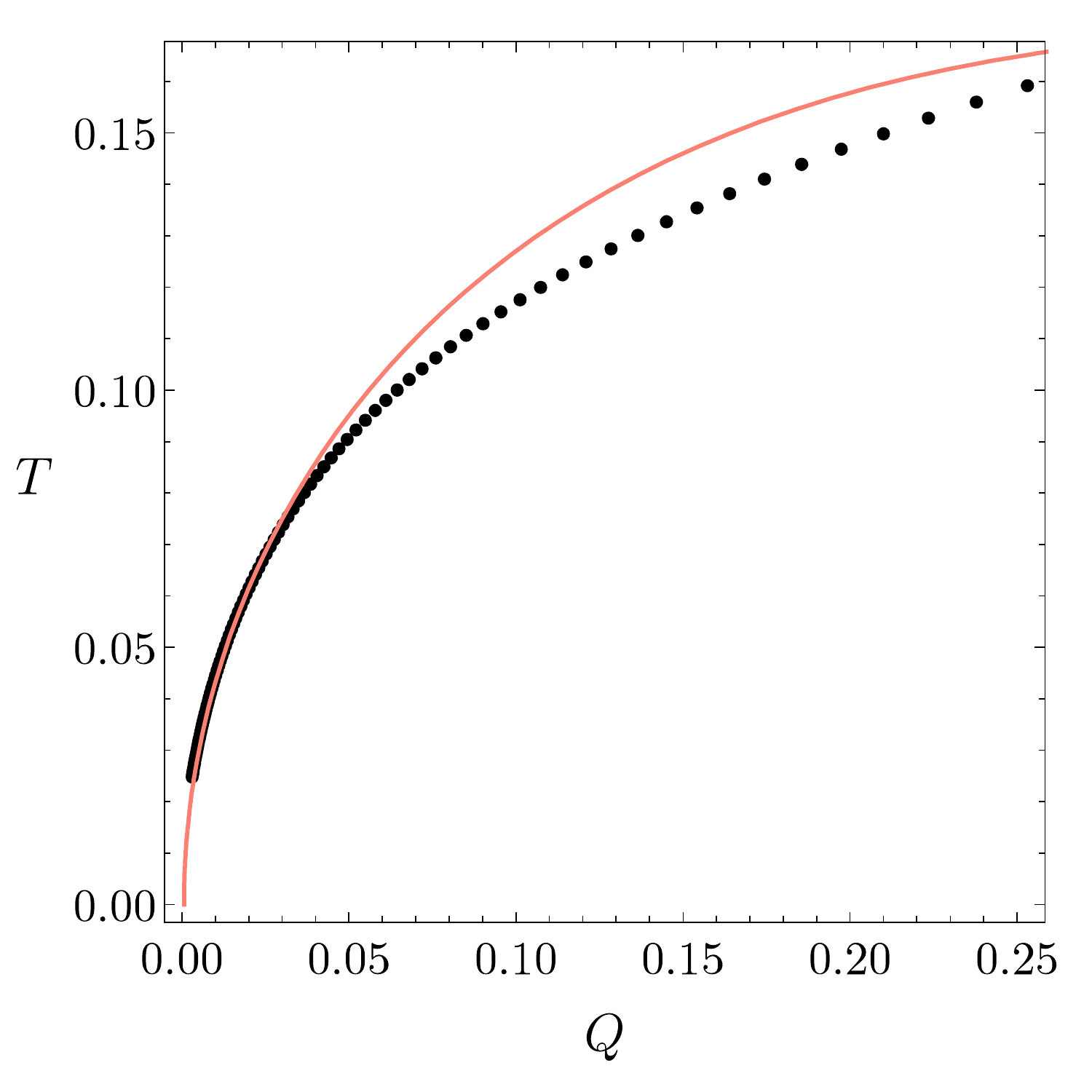}
  \end{minipage}
 \hspace{+2.8em}
    \begin{minipage}[t]{0.36\textwidth}
    \includegraphics[width=\textwidth]{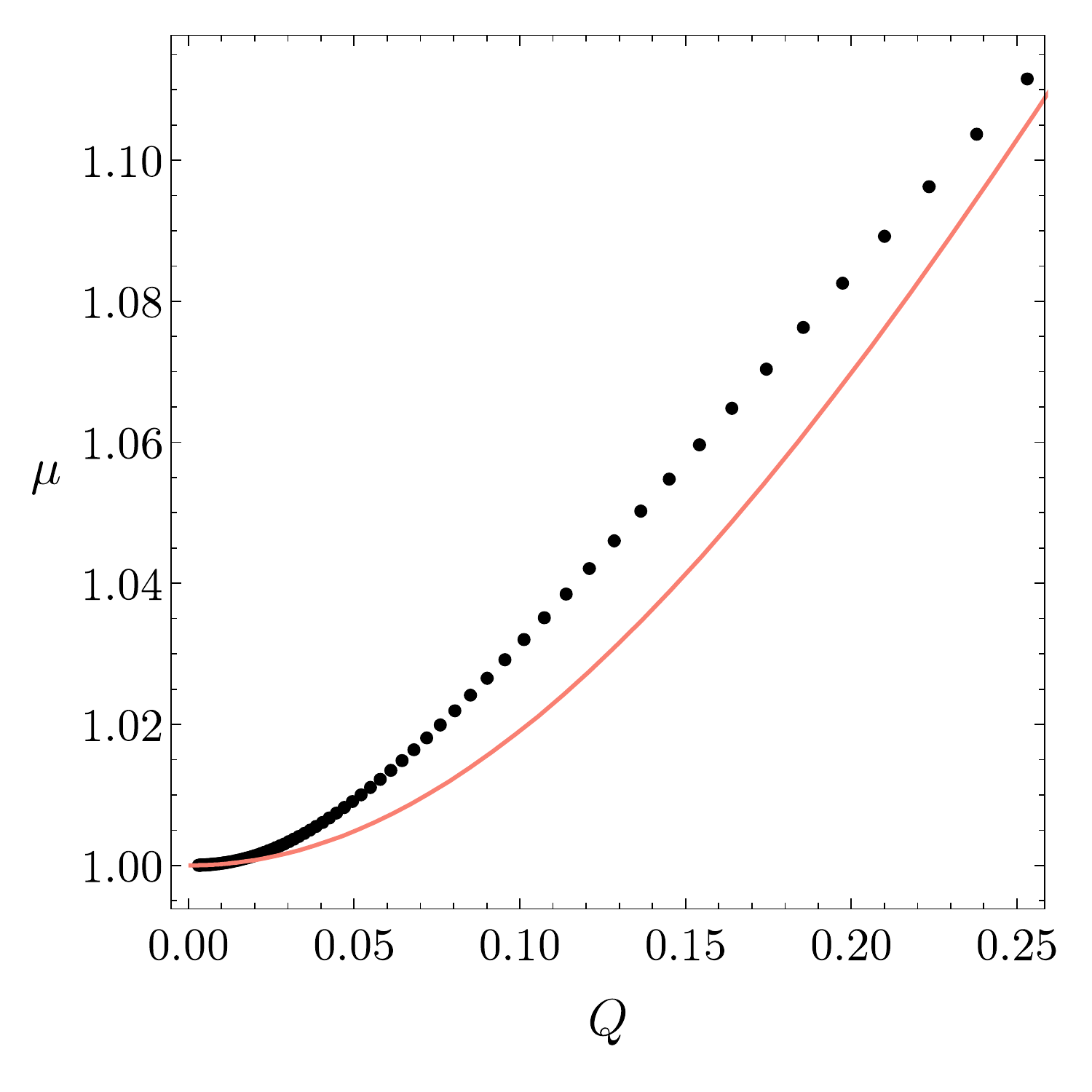} 
  \end{minipage}
      \caption{\label{fig:expncomp} Comparison of the data to the small charge perturbative expansion for the four main thermodynamic quantities. The black disks are the numerical data for the hairy black holes (with $\varepsilon=0.1$) and the red solid line shows the prediction of \cite{Bhattacharyya:2010yg}. As expected, we observe larger deviations in the temperature and chemical potential.}
\end{figure} 
\begin{figure}[b]
\centering
\includegraphics[width=.45\textwidth]{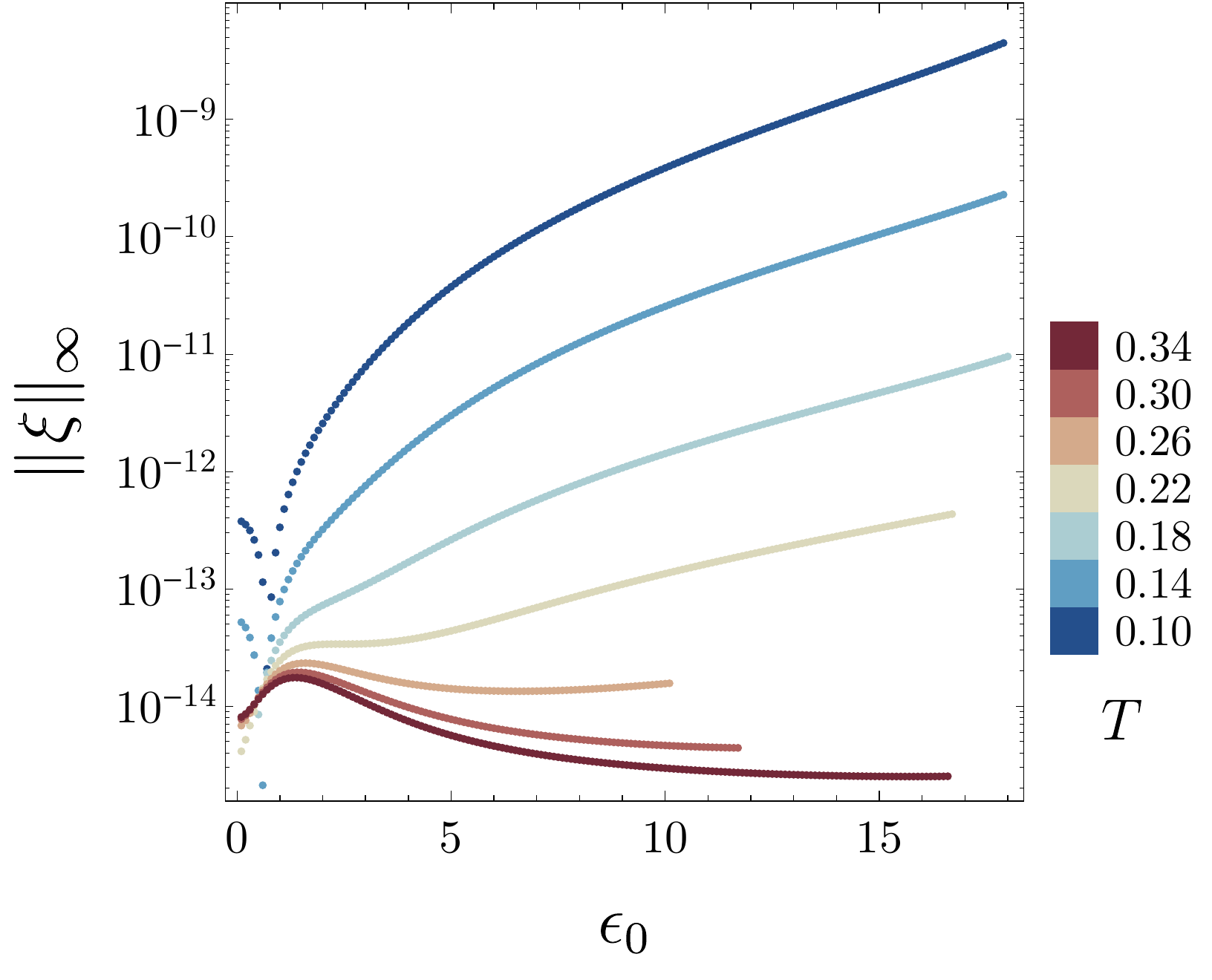}
\caption{The infinity norm of the DeTurck vector across a range temperatures with the number of gridpoints $n=600$. Lower temperature hairy global solutions have the highest norm.}
\label{fig:deturck}
\end{figure}
\subsubsection{Microcanonical ensemble}

Finally the system in which $\delta Q=0$ and $\delta M=0$ is described by the microcanonical ensemble. The preferred phase in this case maximises the entropy. We find that hairy black holes are \emph{only} dominant in this ensemble, see Fig.~\ref{fig:entropy} (and Fig.~\ref{fig:entropyfull} in the Appendix~\ref{sec:therm}). Also in this ensemble $T_2$ plays an important role. In Fig.~\ref{fig:entropy} we plot $S-S_{RN}$ as a function of $M/M_{\mathrm{merger}}$. Here,  $S_{RN}$ corresponds to the entropy of a RNAdS black hole with the same values of $Q$ and $M$ as the hairy solution we are considering, and $M_{\mathrm{merger}}$ to the mass of the RNAdS solution at the onset of the superradiant instability with the same $T$. We see that $S-S_{RN}$ has maximum slope at $M/M_{\mathrm{merger}}=1$, becoming the smallest at $T=T_2$, and increasing again for $T>T_2$. This is a simple consequence of the first law of thermodynamics.

\section{\label{sec:comparison}Comparison with perturbative results}

In this section we compare our numerical results with the perturbative expansion of the hairy black hole solutions of \cite{Bhattacharyya:2010yg}, which are only valid at small asymptotic charges. In \cite{Bhattacharyya:2010yg} the mass and charge are given to sixth order in $\mathcal{O}(\varepsilon^6,R^6,\varepsilon^2 R^4,\varepsilon^4R^2)$, however, the chemical potential is only given to $\mathcal{O}(R^4,\varepsilon^2 R^2,\varepsilon^4)$ and the temperature to $\mathcal{O}(R^3,\varepsilon^2 R^3,\varepsilon^4R)$. In Fig.~(\ref{fig:expncomp}) we present a detailed comparison between our numerical solutions, represented by the black disks, and the expansion of \cite{Bhattacharyya:2010yg} represented by the red solid line. Since the chemical potential and temperature are only determined up to a lower order than the energy, we expect a worse agreement with the numerical data. This is indeed what we observe in Fig.~(\ref{fig:expncomp}). Nevertheless, the observed agreement between the numerical data and the analytic expansion of \cite{Bhattacharyya:2010yg} is reassuring.
  
\section{\label{sec:discussion}Summary and outlook}

In this paper we have studied charged hairy black hole solutions in global AdS$_5$ spacetime using numerical methods. The action that yields these new black hole solutions arises from a consistent truncation of IIB string theory on AdS$_5\times S^5$.  We provided strong numerical evidence that the black hole solutions with the scalar condensate exist between the onset of the superradiant instability and the BPS limit for all values of the hairy black hole charge.

We obtain the smooth horizonless soliton with $\alpha=0$ in the limit $T\rightarrow 0$, while the singular soliton with $\alpha=1$ is reached for any isothermal with $T\neq0$ in the limit where $\epsilon_0$ (scalar field evaluated at the horizon) becomes infinitely large. The fact that these new solutions extend all the way to the BPS limit makes them interesting from the field theory perspective. In fact, from the field theory it is natural that solutions with mass and charge arbitrarily close to the BPS bound should exist, and yet the RNAdS black hole does not saturate such a bound. It is thus reassuring that we did find solutions that saturate the BPS bound; that they turn out to be hairy solutions could have not been anticipated.

We identify the temperature range $T_1<T<T_2$ for which $(\partial M/\partial Q)|_T$ diverges and is marked by complex thermodynamic properties. Globally, we find that the hairy solutions, when they exist, are the preferred phase in the microcanonical ensemble, however, they are subdominant in the canonical and grand-canonical ensembles. In the canonical ensemble, the hairy black holes dominate over RNAdS black holes at low temperatures and become subdominant at high temperatures, however, the hairy solutions are never preferred over pure AdS. Finally, in the grand-canonical ensemble, the RNAdS black holes always dominate over the hairy solutions. These results are recovered in the planar limit.
 
A natural extension of this work is the inclusion of rotation in our setup. Following \cite{Dias:2011at} we started with an equally-rotating Meyers-Perry-AdS$_5$ \cite{1986AnPhy.172..304M,Hawking:1998kw} \emph{ansatz} and constructed a sample of rotating, charged hairy black hole solutions. In this case the hairy black holes moduli space is governed by three parameters $\epsilon_0, y_+$ and $\omega$ where the latter is the black hole angular velocity. For this system it is known that there exist a one parameter family of supersymmetric asymptotically AdS$_5$ black holes \cite{Gutowski:2004ez} with zero scalar field. However, we were unable to obtain supersymmetric hairy black holes. We are exploring the hairy black hole and soliton solution moduli space in greater detail and the results will be presented in a follow up paper~\cite{Markeviciute:2016}.
 
This setup can also be used to analyse other consistent truncations, for instance, consistent truncations of the holographic dual on AdS$_4\times S^7$. In this case the existence of hairy supersymmetric solutions is known (e.g. \cite{Cacciatori:2009iz}) and it would be interesting to explore how non extremal configurations approach these supersymmetric hairy solutions. 
 
\acknowledgments
JM is supported by an STFC studentship. We would like to thank Toby~Crisford and \'Oscar~J.~Dias for reading an earlier version of this manuscript.

\appendix
\section{Numerical validity}
\label{sec:numval}

\begin{figure}[t]
\centering
  \begin{minipage}[t]{0.45\textwidth}
    \includegraphics[width=\textwidth]{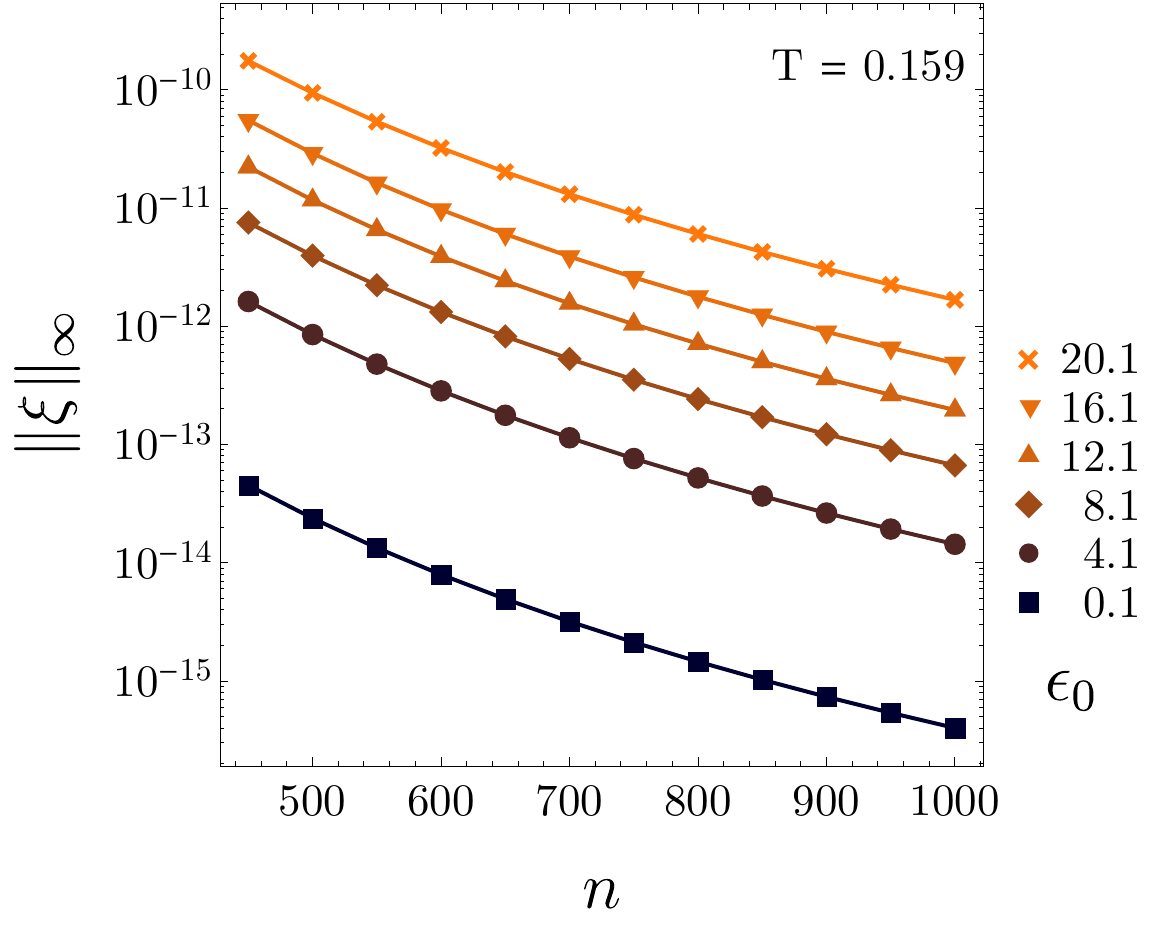}
  \end{minipage}
 \hfill
  \begin{minipage}[t]{0.45\textwidth}
    \includegraphics[width=\textwidth]{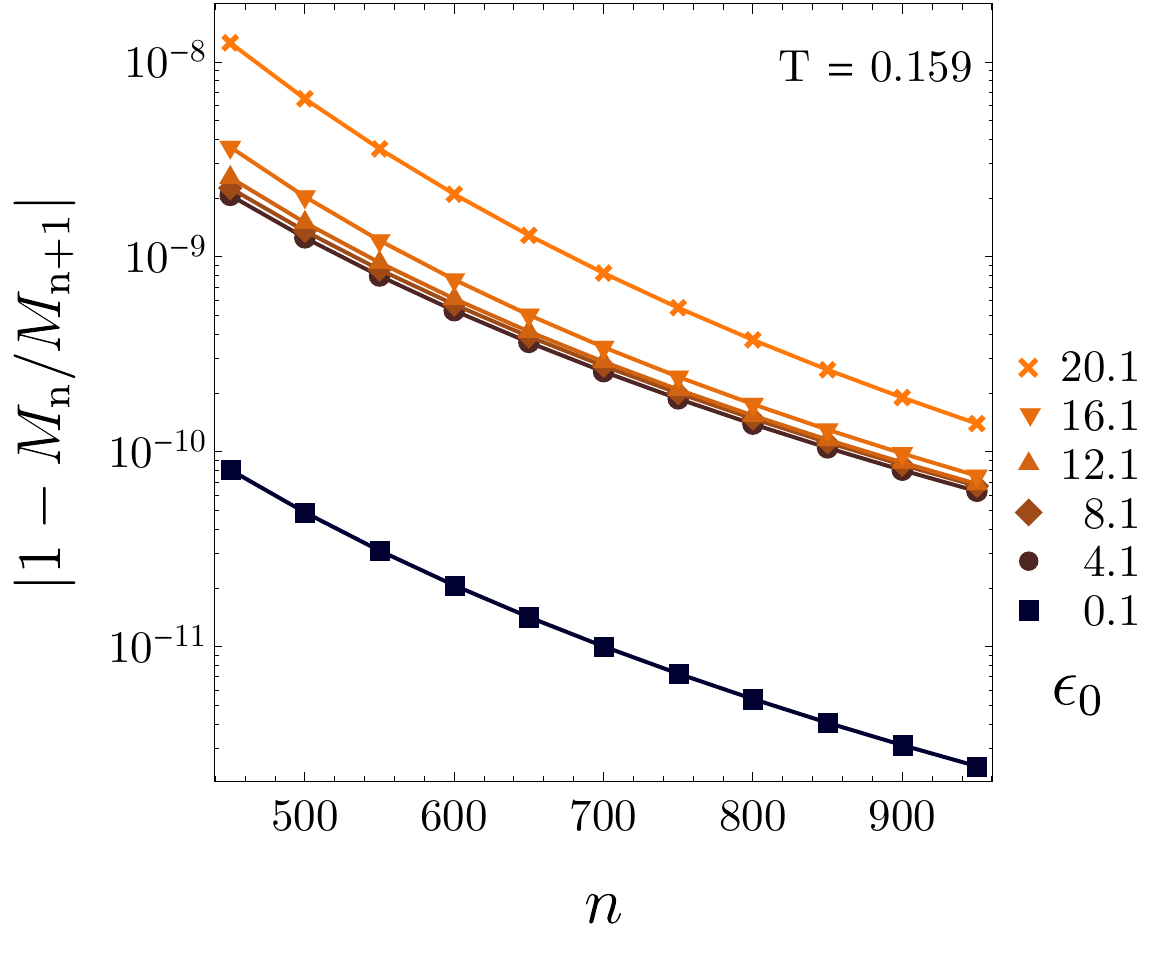}
  \end{minipage}
        \caption{Convergence for the hairy global solutions for different values of the central scalar field density $\epsilon_0$ for one particular temperature. Convergence for the first \emph{ansatz} is at least few orders of magnitude worse. \textit{Left}: The norm of the DeTurck vector versus the grid size. \textit{Right}: Hairy black hole mass error versus the grid size. As the mass involves second derivatives other thermodynamical quantities have at least two order of magnitude better convergence. For $\epsilon_0>20$ convergence falls rapidly.}
    \label{fig:hairyconv}
\end{figure}

\begin{figure}[t]
\centering
  \begin{minipage}[t]{0.45\textwidth}
    \includegraphics[width=\textwidth]{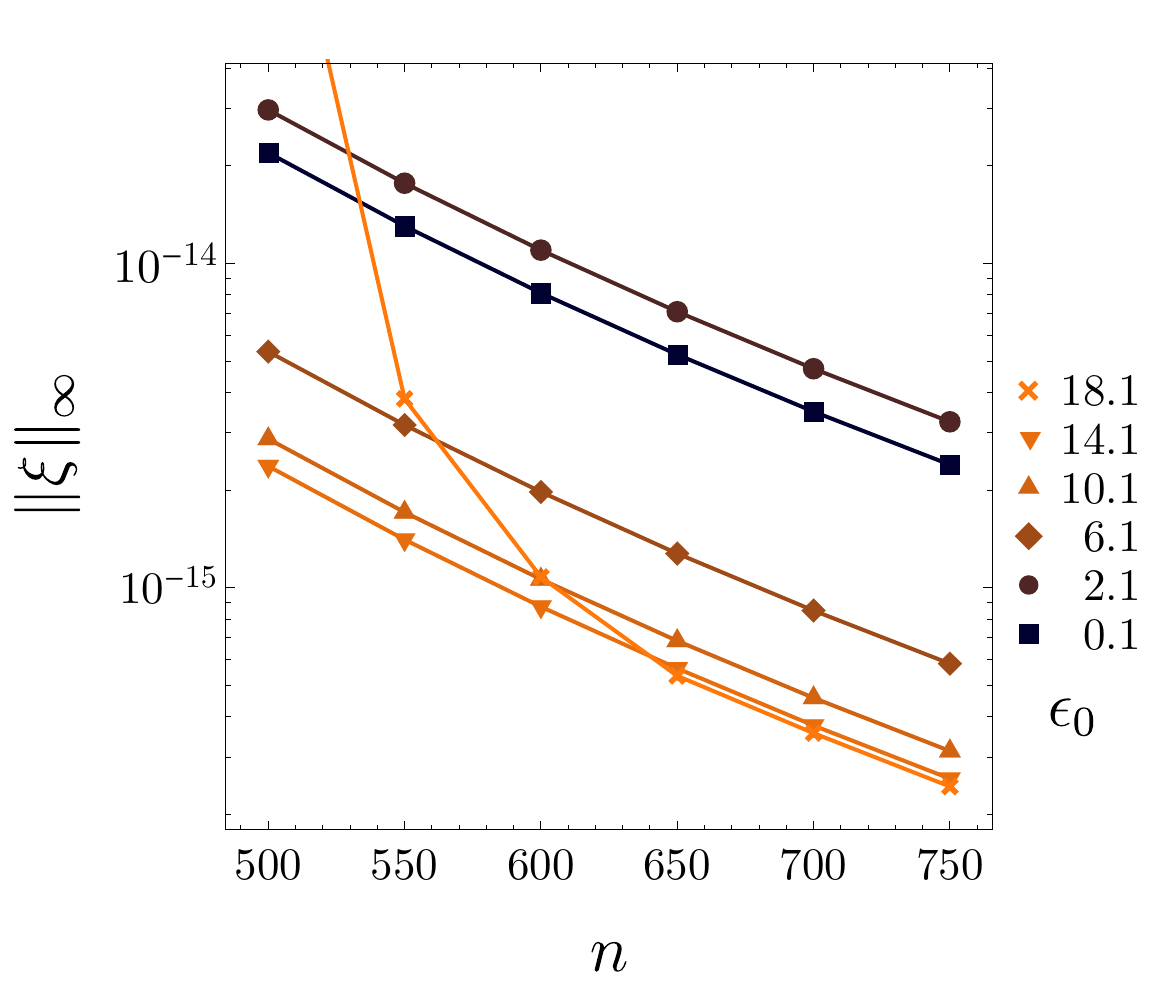}
  \end{minipage}
 \hfill
  \begin{minipage}[t]{0.45\textwidth}
    \includegraphics[width=\textwidth]{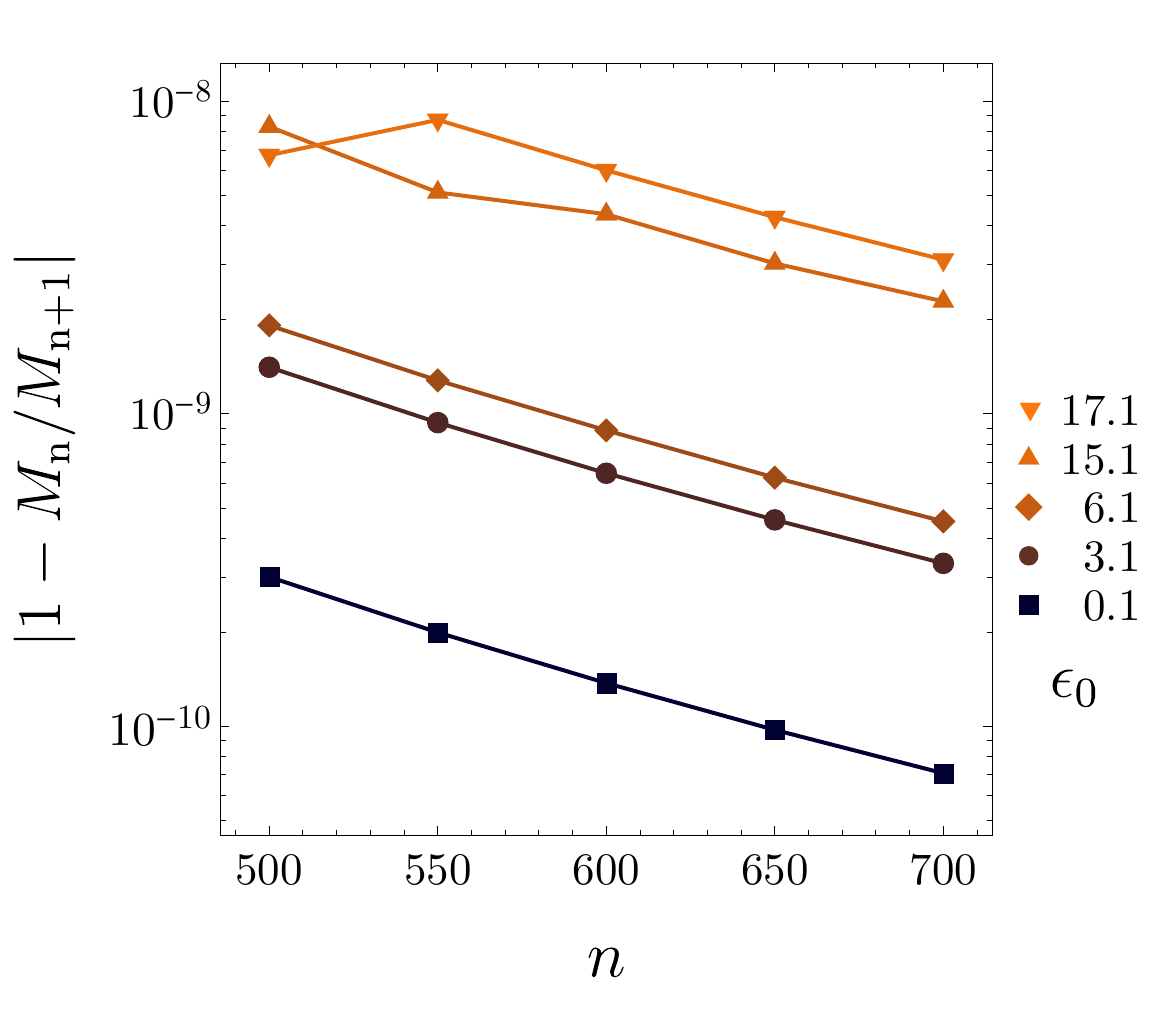}
  \end{minipage}
      \caption{Convergence for the hairy planar solutions for different values of the central scalar field density $\epsilon_0$. As expected it is much better than for the hairy solutions.}
    \label{fig:planarconv}
\end{figure}

We verify that our solutions satisfy $\xi^\mu=0$ to sufficient precision, \emph{i.e.} that our Einstein-DeTurck solutions are also Einstein (see Fig.~\ref{fig:deturck}). We find that low temperature hairy black holes have the highest $\xi$ norm. The pseudospectral methods guarantee exponential convergence with an increasing grid size and we check that all our physical quantities and the norm of the DeTurck vector have this property (Fig.~\ref{fig:hairyconv}, \ref{fig:planarconv}).
 
\begin{figure}[!tp]
\centering
  \begin{minipage}[t]{0.45\textwidth}
    \includegraphics[width=\textwidth]{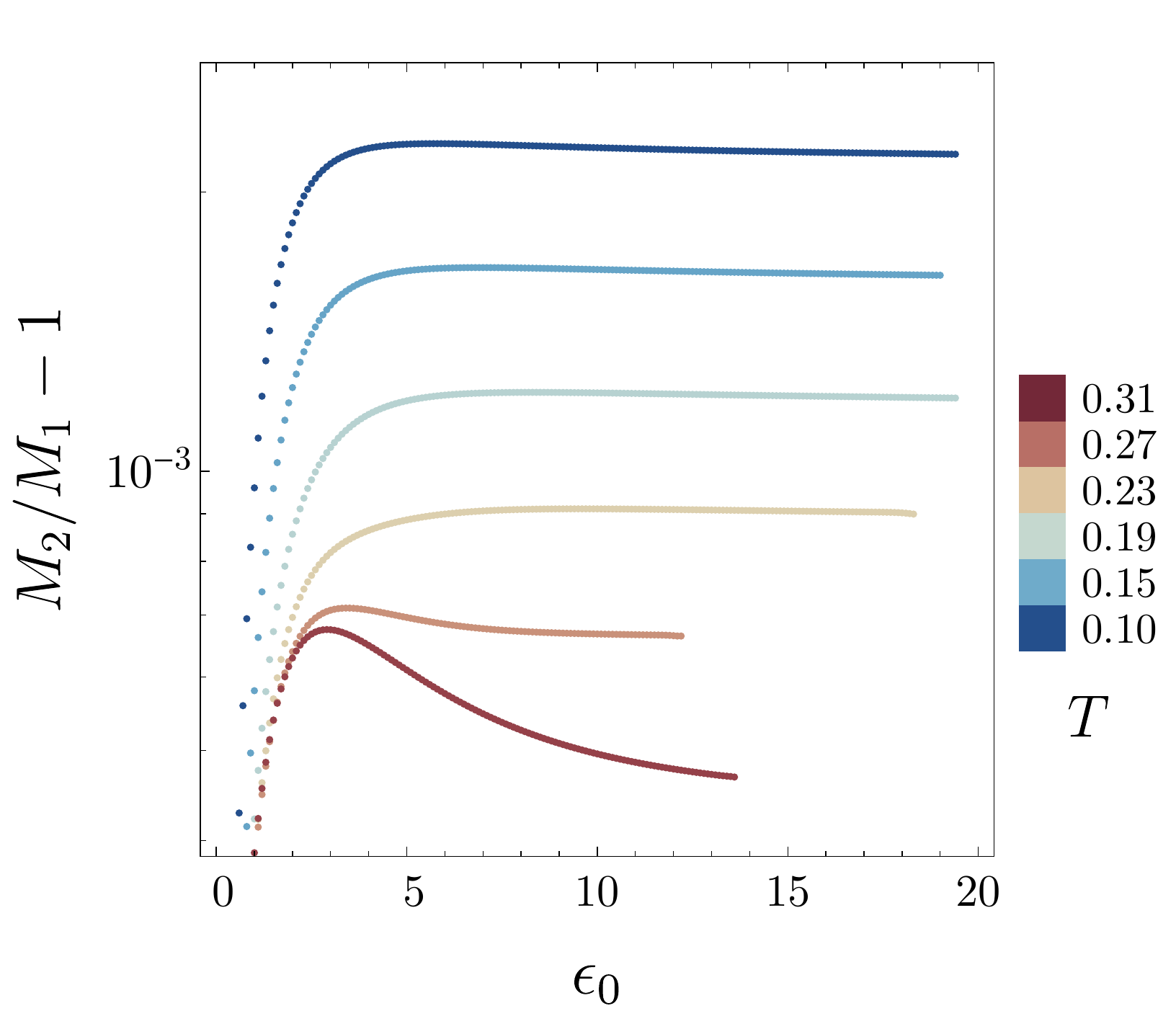}
  \end{minipage}
  \hfill
  \begin{minipage}[t]{0.45\textwidth}
    \includegraphics[width=\textwidth]{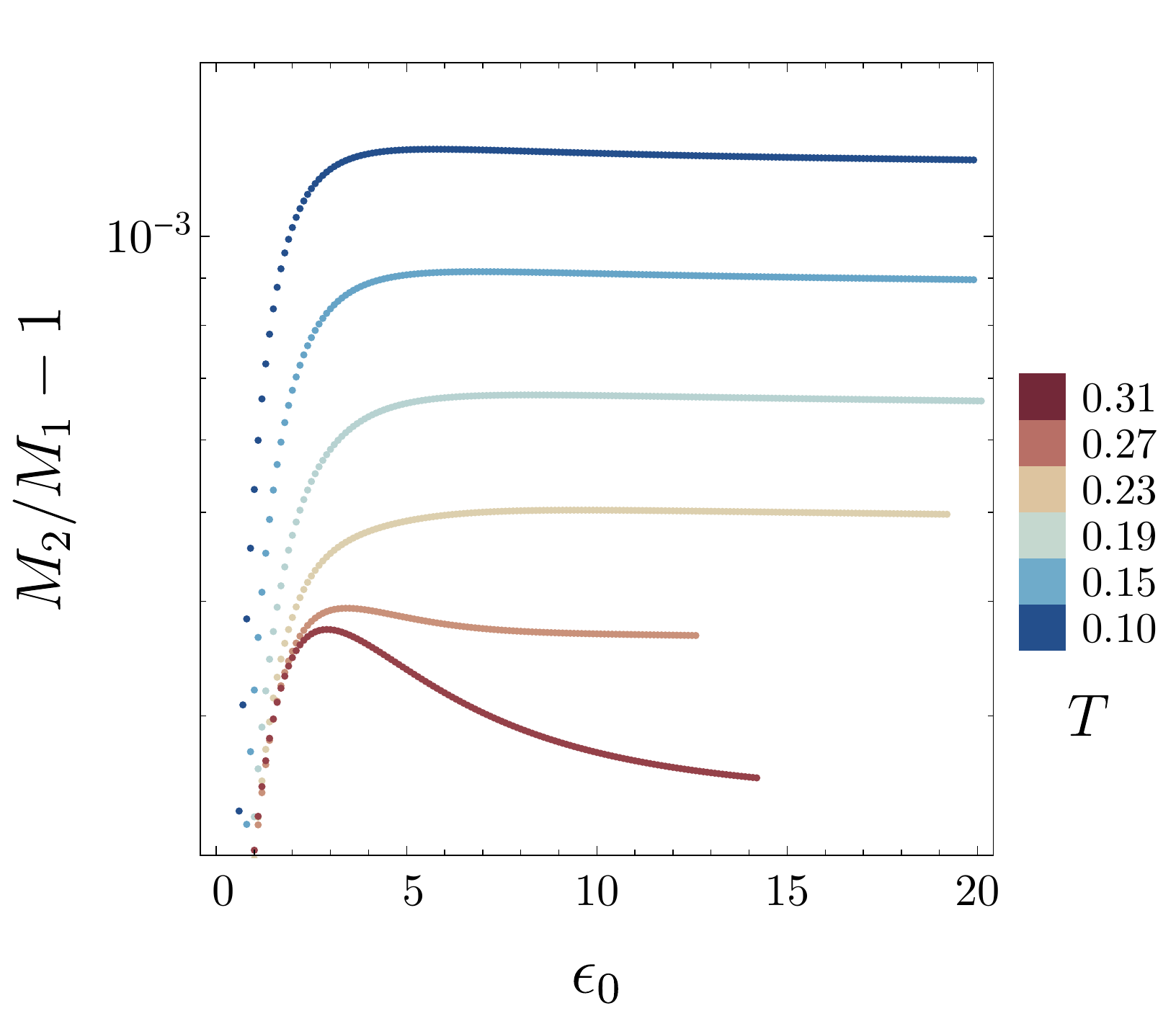}
  \end{minipage}
      \caption{Comparison of the two \emph{ansatz}. $M_1$ is the~\eqref{eq:metric1} \emph{ansatz}, $M_2$ is the~\eqref{eq:metric2} \emph{ansatz}. \textit{Left}: $n=400$ data. \textit{Right}: $n=600$ data. The highest temperature solutions agree the best and the agreement gets worse as we lower the temperature.}
    \label{fig:ansatz} 
\end{figure}
 
\clearpage

\section{Additional figures}
\label{sec:therm}
\begin{figure}[h] 
\centering
\includegraphics[width=1\textwidth]{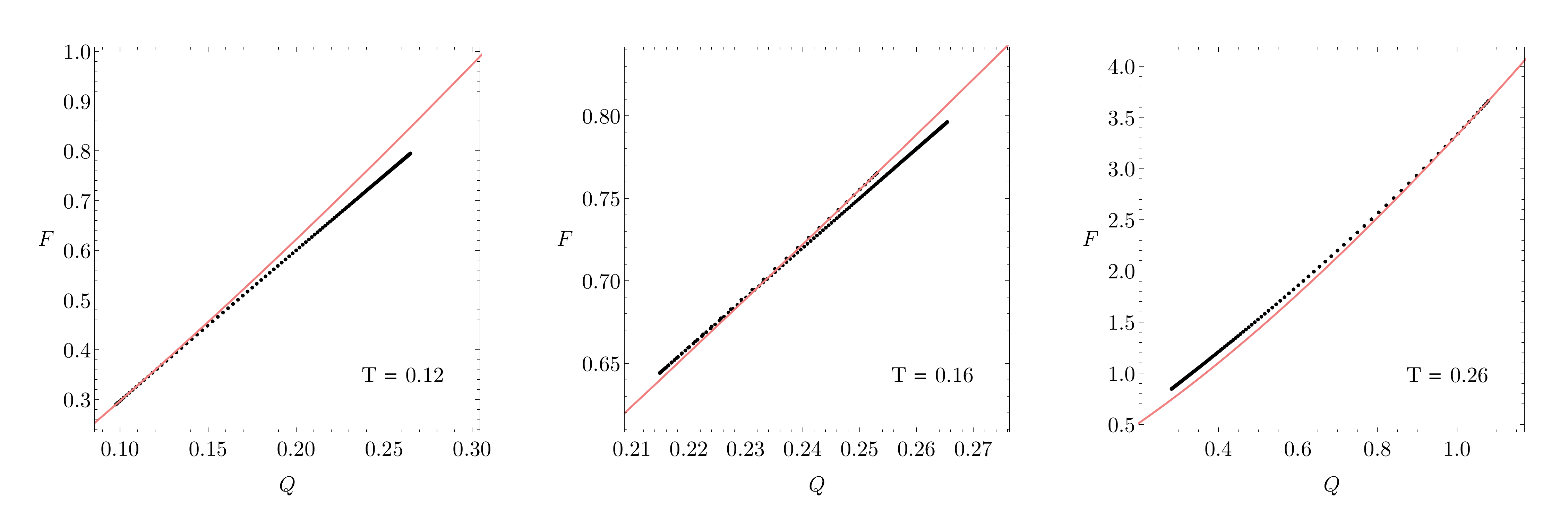}
\caption{\label{fig:helmfull}Canonical free energy versus charge for the hairy solutions for three different temperatures. Red solid line is the corresponding RNAdS solution. For $T<T_1$, (left panel) the hairy solutions dominate over the RNAdS black hole. For $T_1<T<T_2$, the three solutions coexist: two hairy black holes and RNAdS. One of the hairy solutions dominates over the RNAdS, while the other is subdominant (middle panel). For $T>T_2$, the RNAdS black hole is always dominant in the canonical ensemble.}
\end{figure}

\begin{figure}[h]
\centering
  \begin{minipage}[t]{0.33\textwidth}
    \includegraphics[width=\textwidth]{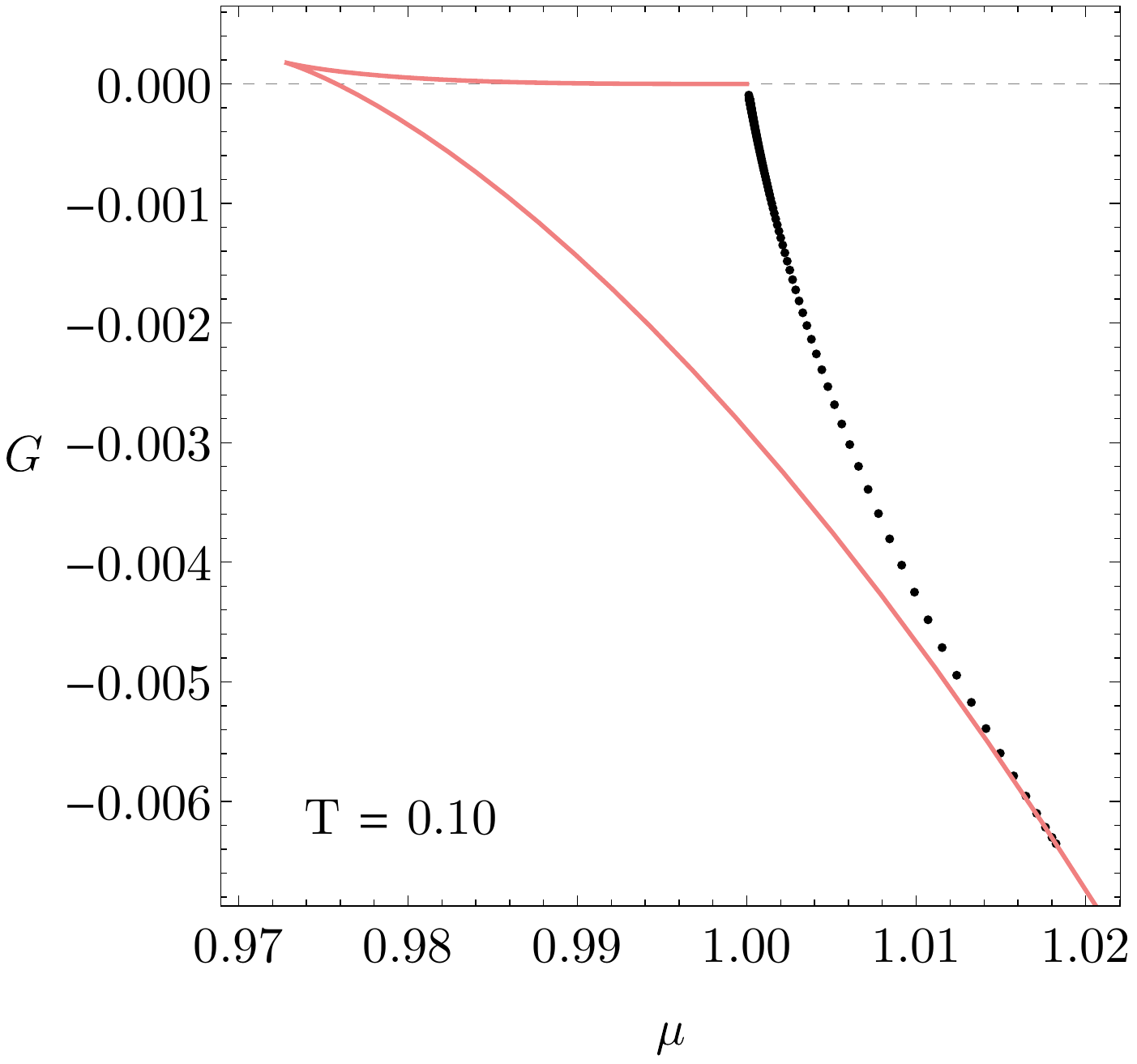}
  \end{minipage}
  \hspace{+5em}
   \begin{minipage}[t]{0.317\textwidth}
    \includegraphics[width=\textwidth]{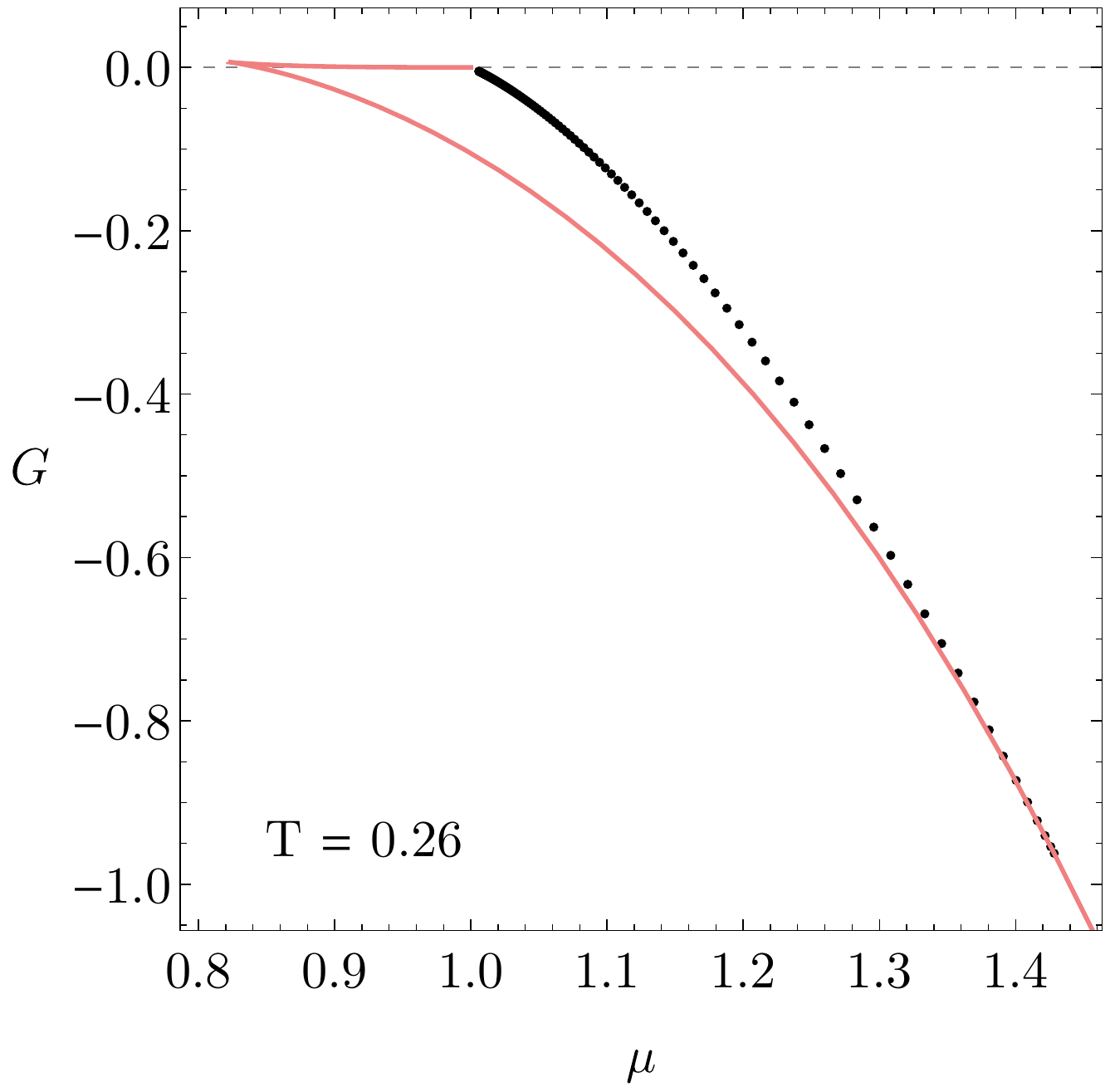}
  \end{minipage}
\caption{\label{fig:gibbsfull} Gibbs free energy versus chemical potential for the hairy global black holes. Red solid line is the corresponding RNAdS solution. In the Gibbs ensemble, the hairy solutions are always subdominant with respect to the RNAdS black hole with the same temperature and chemical potential.}
\end{figure}

\begin{figure}[h]
\centering
\includegraphics[width=1\textwidth]{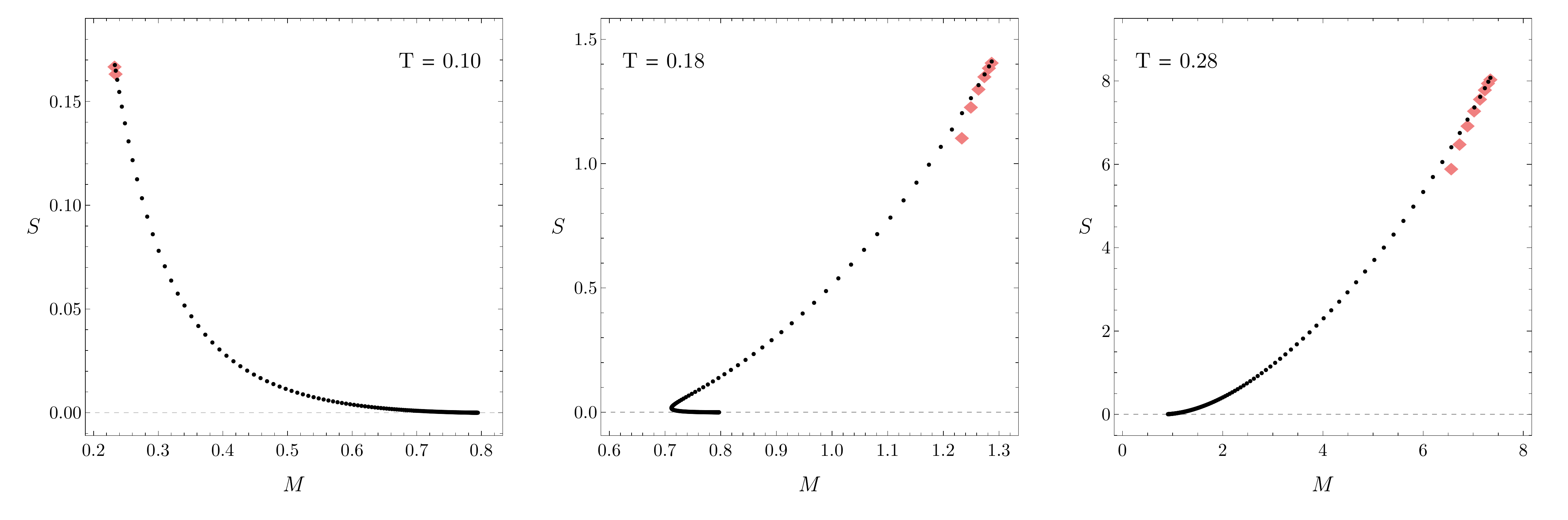}
\caption{\label{fig:entropyfull} Entropy versus mass for the hairy solutions for three different temperatures. Red diamonds correspond to the RNAdS black hole with the same charge. The hairy solutions have higher entropy than the corresponding RNAdS black hole.} 
\label{fig:entropies}
\end{figure}

\begin{figure}[!h]  
\centering
  \begin{minipage}[t]{0.32\textwidth}
    \includegraphics[width=\textwidth]{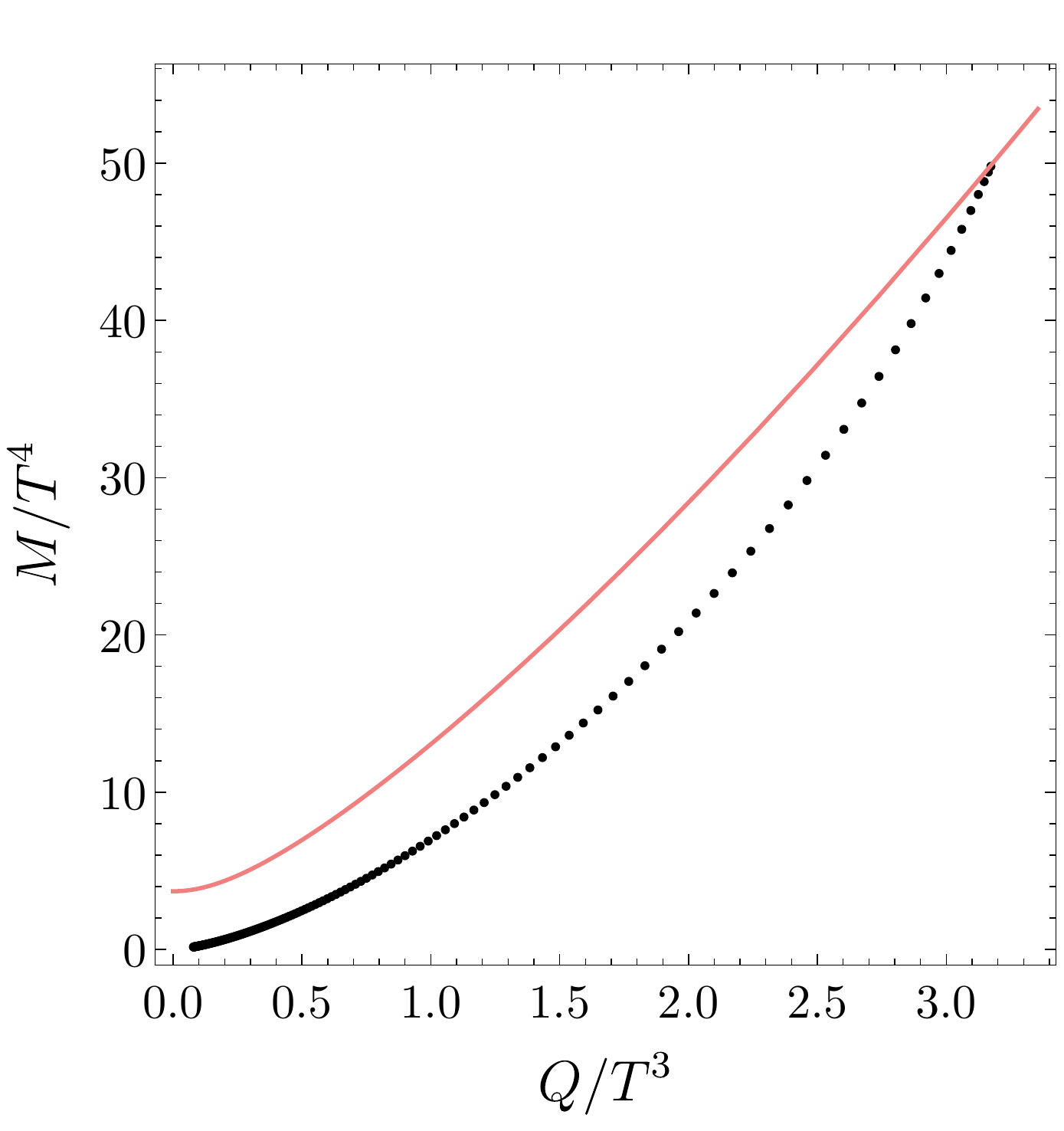}
  \end{minipage}
      \caption{Microcanonical diagram for the hairy black branes. The red line is the family of planar RNAdS black holes: hairy solutions always dominate over RNAdS black holes.}
    \label{fig:pqm}
\end{figure}
\clearpage
\begin{figure}[h]
\centering
  \begin{minipage}[t]{0.30\textwidth}
    \includegraphics[width=\textwidth]{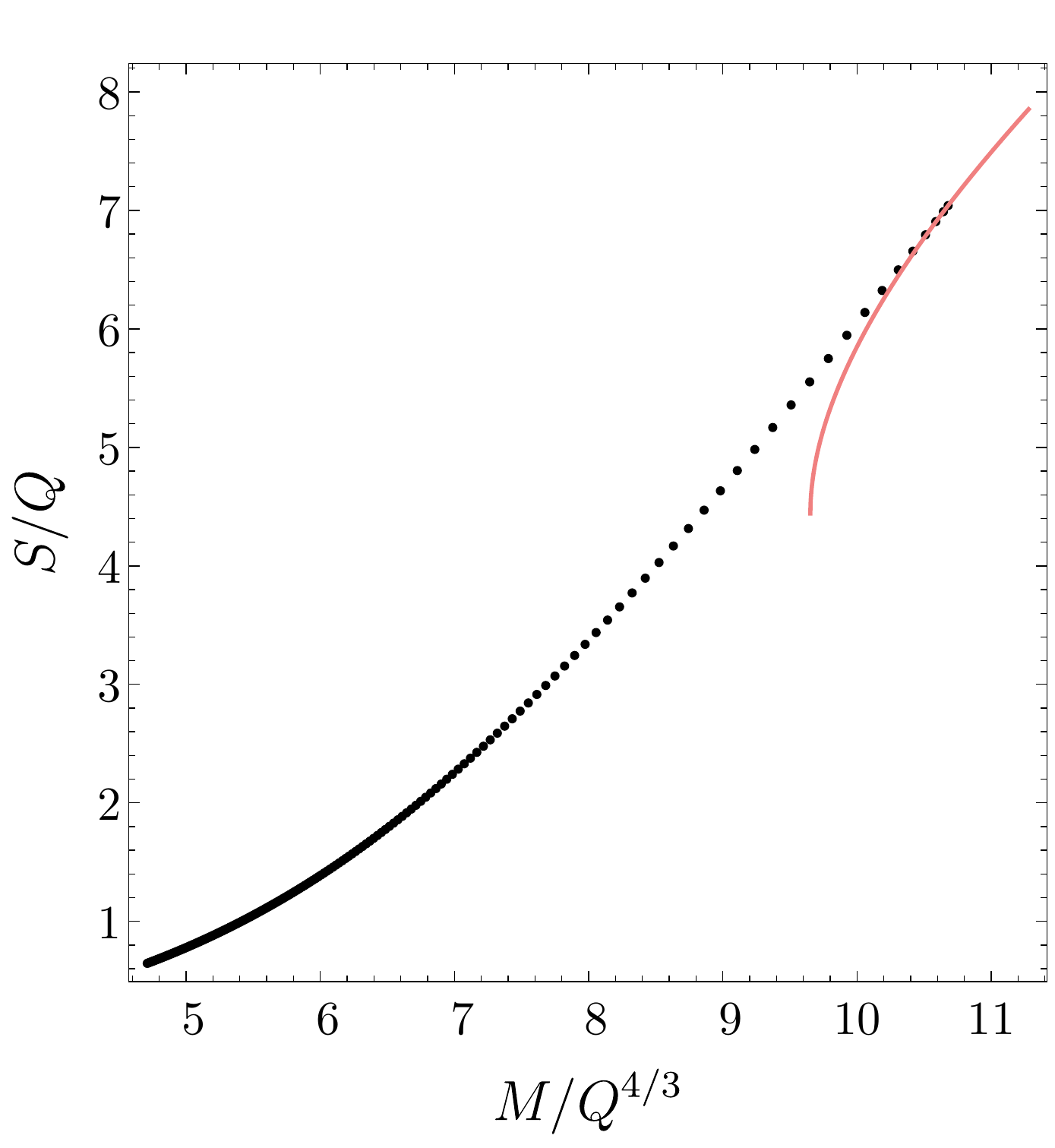}
  \end{minipage}
 \hfill
   \begin{minipage}[t]{0.30\textwidth}
    \includegraphics[width=\textwidth]{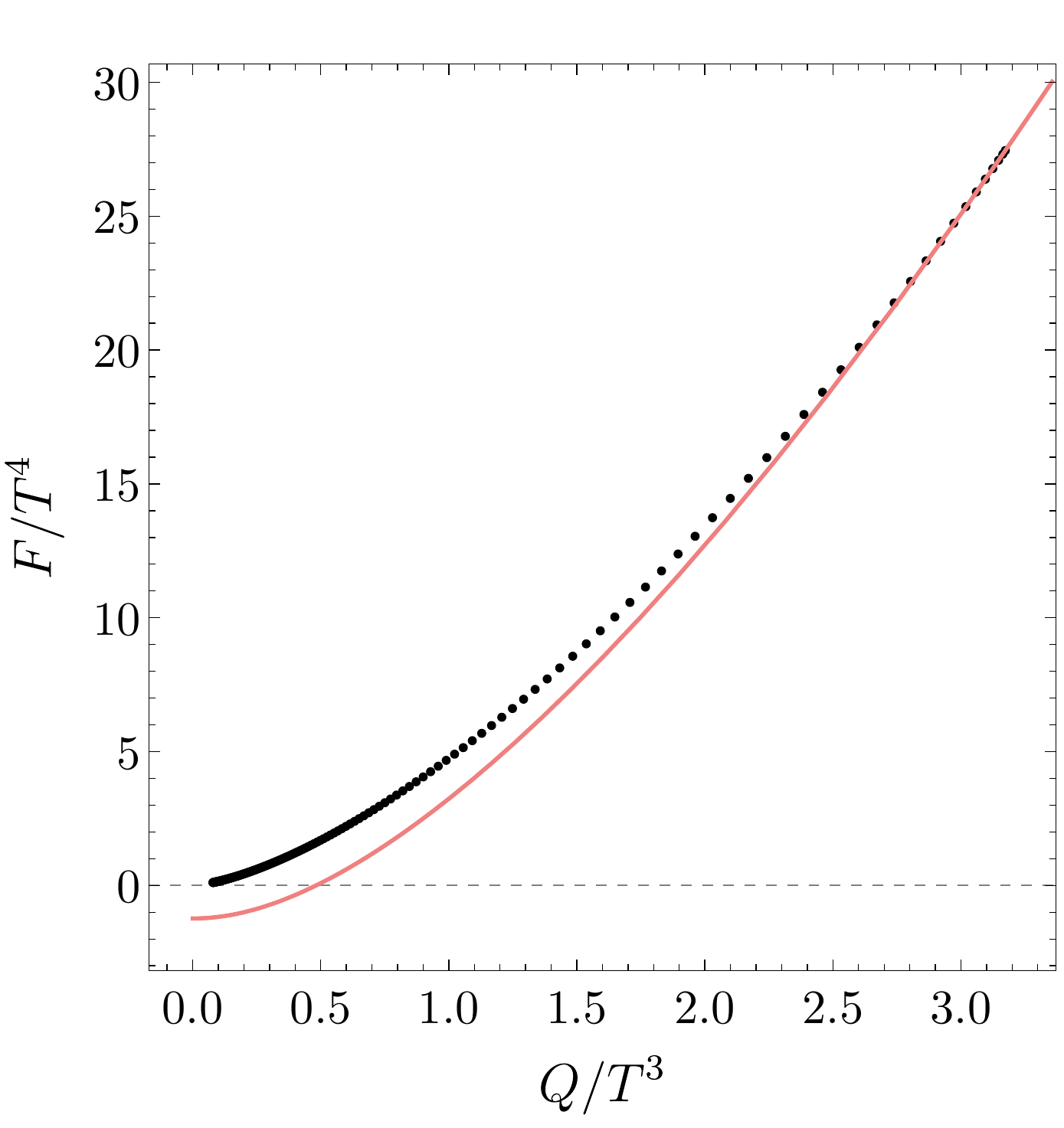}
  \end{minipage}
  \hfill
  \begin{minipage}[t]{0.32\textwidth}
    \includegraphics[width=\textwidth]{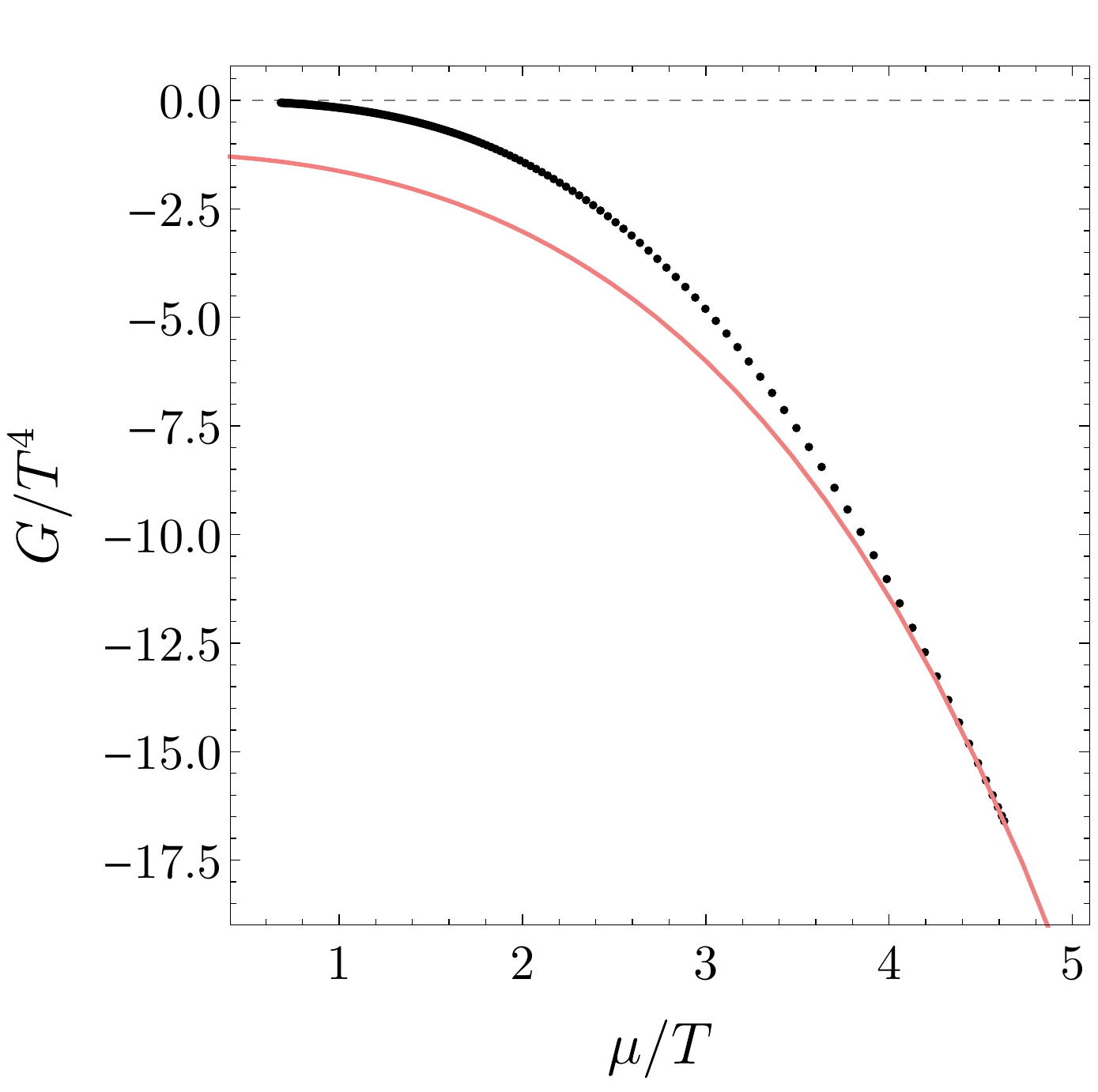}
  \end{minipage}
\caption{\label{fig:planartherm} Scaled thermodynamic potentials for the planar hairy black holes. The red line in each figure is the RNAdS solution in the corresponding ensemble (left to right: microcanonical, canonical and grand-canonical ensembles). The hairy black branes are only dominant in the microcanonical ensemble.}
\end{figure}


\bibliography{hairybh}{}
\bibliographystyle{JHEP}
\end{document}